\documentclass[graybox,usenatbib]{svmult}

\usepackage{mathptmx}       
\usepackage{helvet}         
\usepackage{courier}        
\usepackage{type1cm}        
\usepackage{amsmath}         
\usepackage{makeidx}         
\usepackage{graphicx}        
\usepackage{multicol}        
\usepackage[bottom]{footmisc}
\usepackage{natbib}         
\usepackage[breaklinks,colorlinks,urlcolor=blue,citecolor=blue,linkcolor=blue]{hyperref} 

\makeindex             

\begin{document}

\title*{Observational Techniques With Transiting Exoplanetary Atmospheres}
\author{David K. Sing}
\institute{David K. Sing \at Astrophysics Group, School of Physics, University of Exeter, Stocker Road, Exeter, EX4 4QL, UK, \email{sing@astro.ex.ac.uk}}
%
%
\maketitle
\texttt{Transiting exoplanets provide detailed access to their atmospheres, as
the planet's signal can be effectively separated from that of its host
star.  For transiting exoplanets three fundamental atmospheric
measurements are possible: transmission spectra $-$ where atmospheric
absorption features are detected across an exoplanets limb during
transit, emission spectra $-$ where the day-side average emission of the
planet is detected during secondary eclipse events, and phase curves -
where the spectral emission of the planet is mapped globally following
the planet around its orbit.  All of these techniques have been well
proven to provide detailed characterisation information about planets
ranging from super-Earth to Jupiter size.  In this chapter, I present
the overall background, history and methodology of these measurements.
A few of the major science related questions are also
discussed, which range from broad questions about planet formation and
migration, to detailed atmospheric physics questions about how a
planet's atmosphere responds under extreme conditions.  I also discuss
the analysis methods and light-curve fitting techniques that have been
developed to help reach the extreme spectrophotometric accuracies
needed, and how to derive reliable error estimates despite limiting
systematic errors.  As a transmission spectra derived from primary
transit is a unique measurement outside of our solar system, I discuss
its physical interpretation and the underlying degeneracies associated
with the measurement.} 
\abstract*{
Transiting exoplanets provide detailed access to their atmospheres, as
the planet's signal can be effectively separated from that of its host
star.  For transiting exoplanets three fundamental atmospheric
measurements are possible: transmission spectra $-$ where atmospheric
absorption features are detected across an exoplanets limb during
transit, emission spectra $-$ where the day-side average emission of the
planet is detected during secondary eclipse events, and phase curves -
where the spectral emission of the planet is mapped globally following
the planet around its orbit.  All of these techniques have been well
proven to provide detailed characterisation information about planets
ranging from super-Earth to Jupiter size.  In this chapter, I present
the overall background, history and methodology of these measurements.
A few of the major science related questions are also
discussed, which range from broad questions about planet formation and
migration, to detailed atmospheric physics questions about how a
planet's atmosphere responds under extreme conditions.  I also discuss
the analysis methods and light-curve fitting techniques that have been
developed to help reach the extreme spectrophotometric accuracies
needed, and how to derive reliable error estimates despite limiting
systematic errors.  As a transmission spectra derived from primary
transit is a unique measurement outside of our solar system, I discuss
its physical interpretation and the underlying degeneracies associated
with the measurement.}

\section{Background and History of Exoplanet Atmosphere Observations}
\label{sec:1}
Transiting planets are those that pass directly in front of their
parent star as viewed from the Earth.  During these events,
the planet will block out a proportion of the starlight, which can be
detected by time-series photometry.  To be viewed in this privileged
geometry directly passing in front (or behind) its parent star,
transiting planets require a fortuitous orbital alignment with the Earth.  As
such, transiting exoplanets represent only a small fraction of the
total exoplanet population.  However, the fundamental properties such
as the planetary mass and radius which can be determined (in many
cases nearly free from astrophysical assumptions), and the detailed
spectroscopic information that can also be measured make transiting
exoplanets extremely valuable.   

\begin{figure}[b]
\sidecaption
  \centering
\includegraphics[scale=.33]{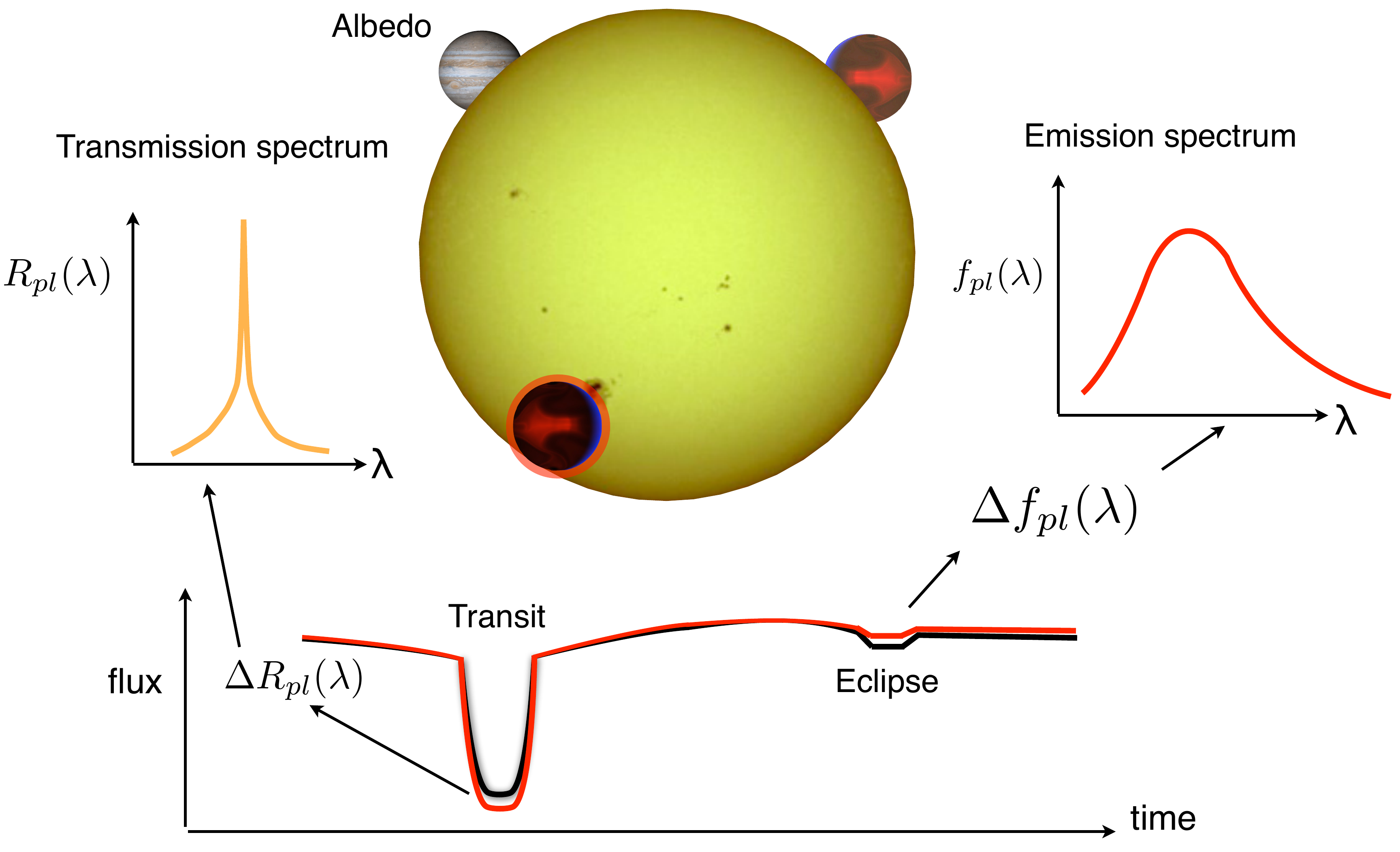}
\caption{Geometry of an exoplanet transit and eclipse event (top
  middle). During the exoplanet's orbit, the light curve of the system is
  monitored, with the transit and eclipse events detected from a drop
  in flux (bottom).  A transmission spectrum is measured by detecting
  a change in transit depth as a function of wavelength
  (top left).  A thermal emission spectrum is measured by detecting a change in the
  eclipse depth as a function of wavelength in the infrared (top right).
  The fraction of reflected light from the planet (the albedo) can be measured by observing the secondary eclipse at optical wavelengths.}
\label{fig:1}       
\end{figure}

The planet-to-star radius ratio can be very precisely measured during
a transit event, as the fractional flux deficit measured from a light
curve, $\frac{\Delta f}{f}$, which is proportional to the projected area between the planet and
star,   
\begin{equation}
\frac{\Delta f}{f} \simeq  \left(\frac{R_{pl}}{R_{star}}\right)^2,
\end{equation}
where $R_{pl}$ and $R_{star}$ are the planet and stellar radii
respectively.  Stellar limb darkening also further modifies the transit
light curve shape.
As the radii of stars can generally be well
determined, the radius of an exoplanet can be measured.  Other
fundamental properties of the exoplanetary system can also be derived
from a transit's light-curve including the planet's inclination, the
semi-major axis of the orbit, and the stellar density.   

For planetary characterisation, the atmosphere is accessible through
the transmission spectrum.  When a planet passes in front of its host
star, some of the starlight will be filtered through the atmosphere at
the planet's terminator and will leave a spectral imprint.  Atoms and
molecules in the atmosphere will absorb and scatter light at
characteristic frequencies, which will make the atmosphere at those
wavelengths opaque at higher altitudes.  In other words, the
atmosphere will be optically thick in slant transit geometry higher in
the atmosphere.  The exoplanet will therefore have a
slightly larger apparent radius at those characteristic wavelengths,
which is directly observable via a deeper transit depth (see Fig.
\ref{fig:1}).  In a sense, a transmission spectrum is essentially an absorption
spectrum, as identification of atomic and molecular species in the
planet's atmosphere are 
identified via absorbed stellar light, though a technically more accurate
description is a transit radius spectrum.  Exoplanet transmission
spectra are now typically constructed by taking time-series
spectrophotometry during a transit event, with the spectra divided
into many wavelength bins in which the chromatic change in transit depth is measured.
At each wavelength bin, a transit
light curve must be fit with a model (typically including instrumental
systematic effects along with a limb-darkened theoretical transit
model) and the planet radius at a particular wavelength, $R_{pl}(\lambda)$, is extracted.
A transmission spectrum is very sensitive to the atmospheric
composition, which is typically among the first bits of information
one learns when a positive signature is identified. 

A planet's atmosphere can also be characterised from an emission
spectrum.  During secondary eclipse, when the planet is seen to pass
behind its host star, the flux contribution from the planet drops to
zero, isolating the flux from the star.  Thus, the light from the
planet can be efficiently separated from that of the star.  The
eclipse depth measured from the fractional flux deficit at secondary
eclipse is directly proportional to the planet-to-star flux ratio,
$F_{pl}/F_{star}$ with,
\begin{equation}
\frac{\Delta f}{f} = \frac{F_{pl}}{F_{star}}\times \left(\frac{R_{pl}}{R_{star}}\right)^2.
\end{equation}
Similarly to a transmission spectra, typically an emission spectra is
constructed by taking time series spectrophotometry, and dividing the
spectra into many different wavelength bins 
in which the eclipse depth is measured and the planet-to-star flux ratio is extracted.
The amount of flux emitted by a planet at infrared wavelengths depends
on its temperature, so this important planetary parameter can be
measured with emission spectra in the infrared. 
Optical secondary eclipses are
sensitive to the reflected light and the geometric albedo of an
exoplanet. 

Finally, when a planet is observed over the course of a full orbit, the flux
contribution from the exoplanet will modulate the total
star-plus-planet flux as its orbital viewing geometry changes, with
atmospheric information obtainable throughout the phase curve.  While
observing an exoplanetary phase curve does not require a transit or
eclipse event, to date most phase curve studies have focused on
transiting exoplanets as they typically offer better overall
constraints.  For instance, the total flux contribution to the phase
curve from the star, which dominates the total signal, can only be
precisely measured during a secondary eclipse event.  
From a phase curve, we can measure the day-to-night temperature
contrast, which informs us about atmospheric recirculation.
Additionally, the abundances of species and atmospheric temperatures
can be mapped around the planet. 

\subsection{A Few Early Results }
\label{sec:1.1}

The following sections are by no means a complete summary or census of
all exoplanet observations relating to atmospheric characterisation
that have been obtained, nor is it meant to be.  Further results can
be found in several review articles 
(e.g. \citealt{2010ARA&A..48..631S, 2014Natur.513..345B,2014PASA...31...43B,2015PASP..127..941C,2017JGRE..122...53D}).  The
intention is to highlight a few representative works to give the
reader a broad introductory overview of the types of atmospheric
measurements that have and can be made regarding transiting exoplanets, and a
flavour of the sort of scientific investigations that result.

\subsection{Transit Observations}
\label{sec:1.1.1}
\begin{figure}[t]
  \centering
\includegraphics[scale=.65]{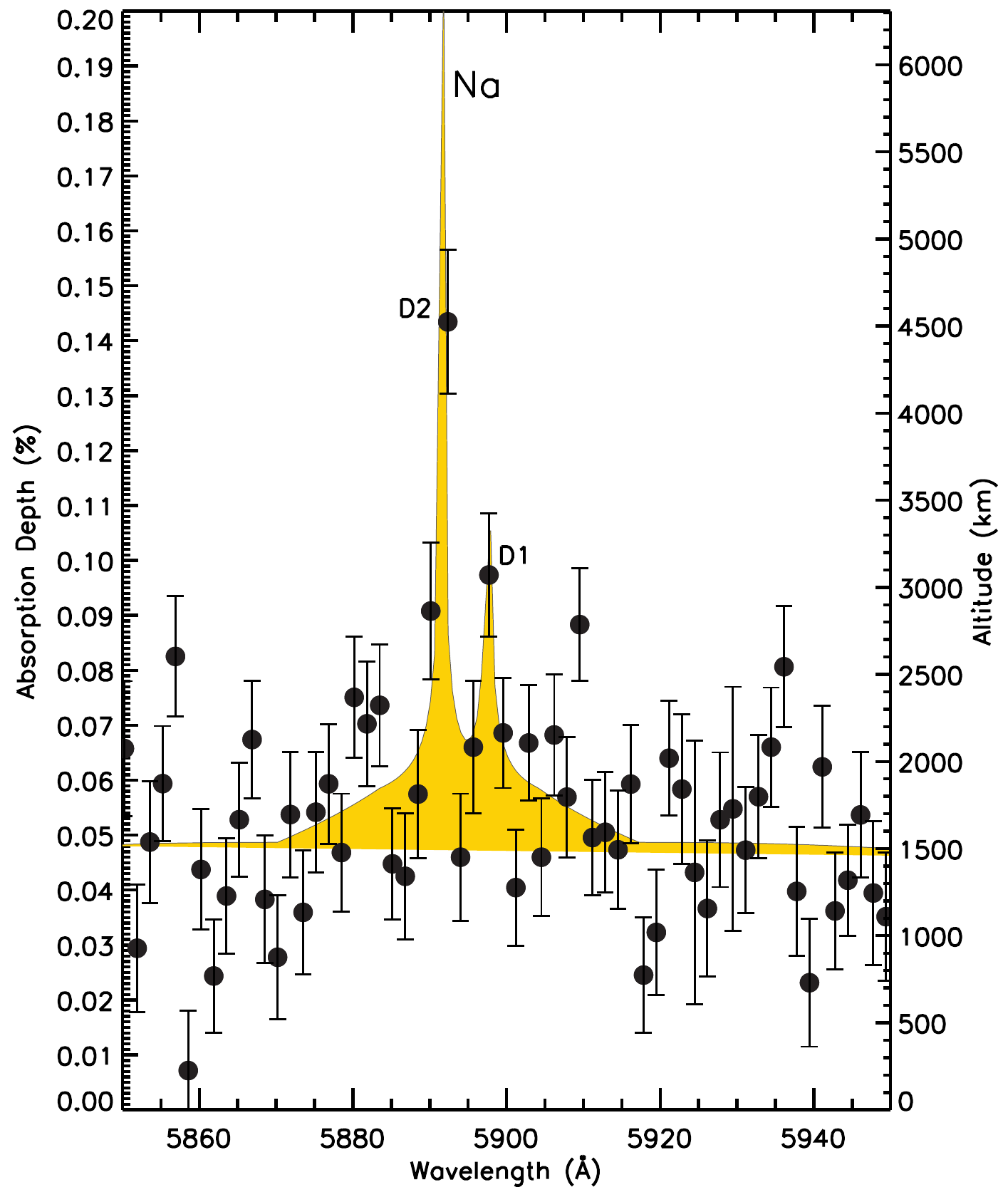}
\caption{Hubble Space Telescope transmission spectrum of HD~209458b showing sodium absorption, which was first detected by \cite{2002ApJ...568..377C}.  Shown is the transmission spectrum adapted from \cite{2008ApJ...686..658S}, which resolves the sodium doublet.  The relative transit absorption depth and relative altitude in the planetary atmosphere are both indicated.}
\label{fig:2}       
\end{figure}

As recently as 2006 there were fewer than 10 transiting exoplanets
known, and the atmospheric characterisation of these planets were
largely limited to a few select cases.  Among the first characterised
planets was HD~209458b, a exoplanet first found using the radial
velocity technique \citep{2000ApJ...532L..55M} which was the first planet
discovered to also transit by \cite{2000ApJ...529L..45C} and \cite{2000ApJ...529L..41H}.  Having initially been discovered by the radial velocity
technique helped make HD~209458b particularly well suited for followup
atmospheric characterisation.  The radial velocity method requires
bright target host stars in order to obtain sufficient signal-to-noise
at high spectral resolution, which has historically limited radial
velocity discoveries on moderately sized telescopes to V-magnitudes
typically brighter than about 10.  Transiting around such bright host
stars opens up the possibility to perform very high precision
photometry, as sufficiently large numbers of photons can be gathered
during the short $\sim$hour long transit durations.  Shortly after the
discovery of transits, the Hubble Space Telescope (HST) observed
HD~209458b during four transits with the Space Telescope Imaging
Spectrograph (STIS).  The resulting transit light curve \citep{2001ApJ...552..699B} was unprecedented in quality with 110 parts-per-million
photometric accuracies with an 80 second cadence.  The high quality demonstrated not only the feasibility to
perform atmospheric studies but also to detect Earth-sized exoplanets
in transit around Sun-like stars (the NASA Kepler mission was selected
12 months later).  With the same STIS data, \cite{2002ApJ...568..377C}
made the very first detection of an exoplanet atmosphere by observing
excess absorption in the Na doublet (See Fig. \ref{fig:2}).  Alkali metal
absorption had been previously predicted to be present in the
atmosphere by \cite{2000ApJ...537..916S}, as the high temperature
allows atomic sodium to exist in the gas phase. 

The HD~209458b sodium detection was a good early indication of the
high photometric precisions which would be needed to regularly make
exoplanet atmosphere detections, as the observed sodium signal
measured by \cite{2002ApJ...568..377C} represented a deeper transit by
just 232 parts-per-million (ppm) relative to the adjacent wavelengths
and precisions of 57 ppm were needed to secure a 4-$\sigma$ confident
detection.

As the transit technique is reliant upon the starlight to make a
measurement, and not light from the
planet itself, a transmission spectrum can be
studied across a much broader wavelength range than traditionally
possible when observing only emitted radiation.  Shortly after the
sodium detection, HST STIS followed HD~209458b up again but this time
in the far-ultraviolet (FUV), targeting the H~I Ly-$\alpha$ line.  An extended exosphere of
H~I was expected around the exoplanet, which could in principle be
detected against the stellar chromospheric Ly-alpha emission during a
transit event.  However, the resulting transit depth as observed in H
I  (15\%, \citealt{2003Natur.422..143V}) was far in excess of the planet
itself or even its Roche lobe, indicating hydrogen atoms are
vigorously escaping from the planet.  Oxygen and carbon were also
detected in the exosphere by STIS shortly thereafter
\citep{2004ApJ...604L..69V}.  With the second transiting planet discovered in 2003
\citep{2003Natur.421..507K}, for a short time, the number of atmospheric
detections made via the transit method outpaced the number of
transiting planets.  Even with thousands of transiting planets known
today, HD~209458b still resides as one of the very best theoretical
targets for atmospheric characterisation.  

Transmission spectral observations from the ground also proved
feasible, first at high resolution.  \cite{2008ApJ...673L..87R} observed
HD~189733b across optical wavelengths with the Hubby-Eberly Telescope
at high resolution (spectral resolution, R $\sim$60,000).  Significant absorption was detected
in-transit in the Na D lines, indicating deeper transit depths by
672$\pm$207 ppm.   Similarly, \cite{2008A&A...487..357S} analysed optical HD
209458b transit data from the Subaru telescope at high resolution
($R\sim$45,000).  A significant non-linearity in the CCD had to be
corrected, and once applied significant Na absorption was also
detected (1350$\pm$170 ppm).  The Na absorption from HD~209458b first
observed with Hubble was not only confirmed, but the Na line profile
itself also matched well between the observations
(\citealt{2002ApJ...568..377C, 2008ApJ...686..658S, 2008A&A...487..357S}).  Historically,
such agreement between observations has not always been case, but
these early results did prove to place exoplanet atmospheric
observations on a solid foundation.  

The shape of absorption line profiles in transmission spectra can be
used to probe different altitudes of exoplanet atmospheres.  The wings
of a line have a lower optical depth, and probe lower, cooler parts of
the atmosphere, while the extended core of the line probes out to much
higher altitudes.  Because the atomic Na D resonance lines are very
strong and have little other absorbers obscuring the view in their
wavelength region, they are ideal for probing a wide altitude-pressure
range.  Their line profiles have been well-measured from HST and the
ground for HD 189733b and HD 209458b (see also \citealt{2012MNRAS.422.2477H, 2011ApJ...743..203J}).
Thermosphere layers were detected in both planets (e.g. \citealt{2011A&A...527A.110V} for HD~209458b
and \citealt{2012MNRAS.422.2477H} for HD~189733b).
The thermosphere is an extended region above the typical lower layers
of the atmosphere (troposphere and stratosphere), but below the
exosphere, where UV radiation is absorbed and temperatures rise as a
function of height. The cores of the Na lines extend to higher
altitudes in the presence of a thermosphere, as the hotter
thermospheric temperatures increase the pressure scale height and puff
up the atmosphere - leading to larger transmission spectral features.
For instance, HARPS observations of HD 189733b have resolved the Na
lines up to an altitude of 12,700 km \citep{2015A&A...577A..62W}.  With
high enough signal-to-noise (S/N), a change in pressure scale height
at different altitudes can be directly detected, which then informs us
about the temperature change in the atmosphere between the upper and
lower layers (see section \ref{sec:1.4} below). 

\subsection{Eclipse Observations}
\label{sec:1.1.2}
\begin{figure}[t]
\sidecaption
  \centering
\includegraphics[scale=.5]{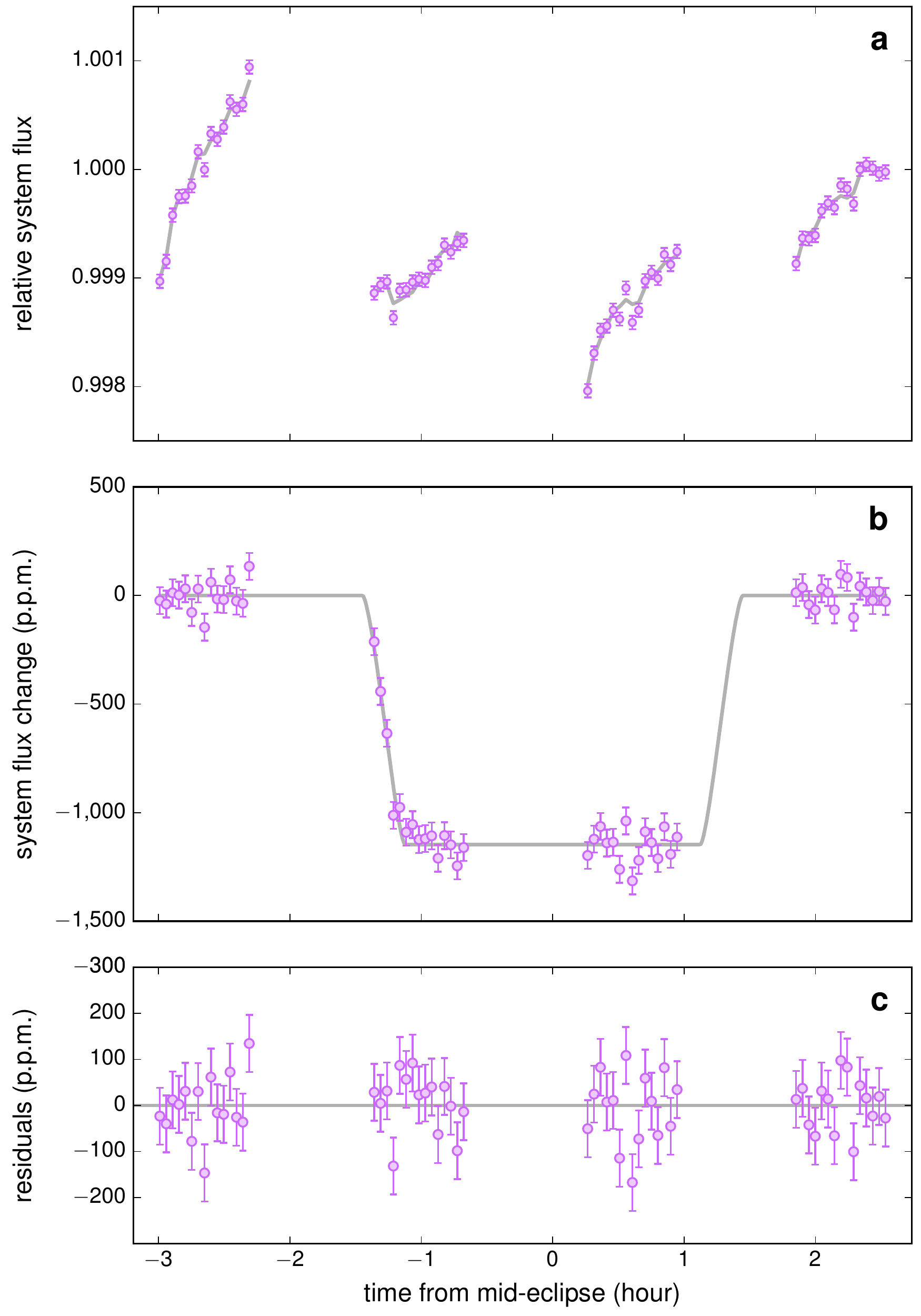}
\caption{Example of a secondary eclipse observation of WASP-121b from
  the Hubble Space Telescope WFC3 (from \citealt{2017Natur.548...58E}).  Plotted (a) is the raw normalized flux with photon noise
error bars and the best-fit eclipse and instrument systematic trend model, (b)
the relative change in system flux after correcting for instrument
systematics, and (c) the residuals between the data and best-fit model
showing precisions of 64 parts-per-millon.}
\label{fig:W121ecl}       
\end{figure}

In August of 2004 the HST STIS instrument failed, preventing further
observations until a repair could be made.  However, a year earlier
the Spitzer Space Telescope was launched, and while it was not
designed for exoplanet transit observations, the first secondary
eclipse measurements were achieved shortly after in 2005 by
\cite{2005Natur.434..740D} for HD~209458b, while 
\cite{2005ApJ...626..523C} targeted TrES-1b.   As Spitzer is an
infrared telescope, it is sensitive to the longer wavelengths needed to
probe a hot exoplanet closer to its $\sim$1000 K black-body emission peak.
Additionally, at these longer wavelengths the parent star is significantly fainter, which further increases the contrast.
Both exoplanets were observed to have
secondary eclipse depths near 0.25\% and from these single eclipse
measurements, brightness temperatures (which assumes the planet
emits as a black body) could be derived.  The timing of the
secondary eclipse measurements placed informative constraints on the
orbital eccentricity.  With these results, Spitzer in effect took the
reins over a period of a few years as the leading (and virtually only)
exoplanet characterising instrument.  The repair of HST and
installation of the WFC3 instrument in 2009 also enabled high-precision secondary eclipse measurements (e.g. Fig. \ref{fig:W121ecl}).

With multiple secondary eclipse measurements at different wavelengths,
Spitzer was able to build up the first rough broadband emission
spectra, which could be compared to theoretical atmospheric models.
An early result came from \cite{2008ApJ...673..526K} who used
Spitzer to build up a spectrum from five photometric channels between 3
and 30 microns for HD~209458b.  These broadband spectra are very low
resolution ($R\sim$5 to 10), thus no specific molecular features can be
resolved, though in principle very large-scale differences between
atmospheric models are detectable.   The HD~209458b measurements
indicated the presence of a stratosphere where the temperature is seen
to rise (rather than fall) at higher altitudes.  Rising temperatures
create conditions where hot atmospheric gas lies above cooler gas,
which gives rise to spectral emission features.  If a planetary
atmosphere has a temperature-pressure (T-P) profile which only cools
off with higher altitudes, only spectral absorption signatures from
the overlying cooler gas could be observed.  For HD~209458b, \cite{2008ApJ...673..526K} 
found the wavelengths between 4 and 10 microns were
significantly higher than would be expected from non-inverted T-P
profiles which indicated H$_2$O was in emission (also see \citealt{2007ApJ...668L.171B}).  While these specific measurements did not generally hold
up to further scrutiny (see section \ref{sec:1.2.2} below), they were a strong
motivation into further sophisticated theoretical investigations of
the atmospheric dynamics and chemistry of highly-irradiated gas giant
planets, and the first steps towards constraining such models with
eclipse observations.   

Atmospheric windows in Earth's own atmosphere also permit exoplanet
transit and eclipse observations from the ground.  However,
the challenges are formidable as precision photometry ($\sim$100 ppm) must
be preformed for several hours against strongly changing weather
conditions and instrument instabilities.  Furthermore, for eclipse
observations the largest signals occur at longer wavelengths which are
generally inaccessible due to telluric H$_2$O absorption.  Thus,
exoplanet eclipse observations have generally been performed in the z',
J, H, and K atmospheric windows.  Theoretical predictions early on
\citep{2007ApJ...667L.191L} predicted the hottest planets ($\sim$2500 K)
would likely have significant and detectable thermal emission in the
red-optical z' band, and shortly after \cite{2009A&A...493L..31S}
detected the thermal emission of Ogle-Tr-56b while at the same time
\cite{2009A&A...493L..35D} detected the secondary eclipse of TrES-3b
in the K band
with the WHT telescope.  These results directly demonstrated a very
large number of exoplanet eclipses would eventually be observable,
given that Ogle-Tr-56b orbits a very distant 16th magnitude star and
the TrES-3b result was made with a modestly sized 4-meter telescope. 
\cite{2011AJ....141...30C} observed eclipses of WASP-12b with the CFHT telescope at J, H, and K band which
further demonstrated high-precision measurements could be obtained
from the ground. 
Thus, exoplanet atmospheric characterisation from eclipses were no
longer confined to flagship space telescopes, and transit and eclipse
observations became demonstrably accessible for a wide range of targets using a
wide variety of instrumentation.  

\begin{figure}[t]
  \centering
\includegraphics[scale=.45]{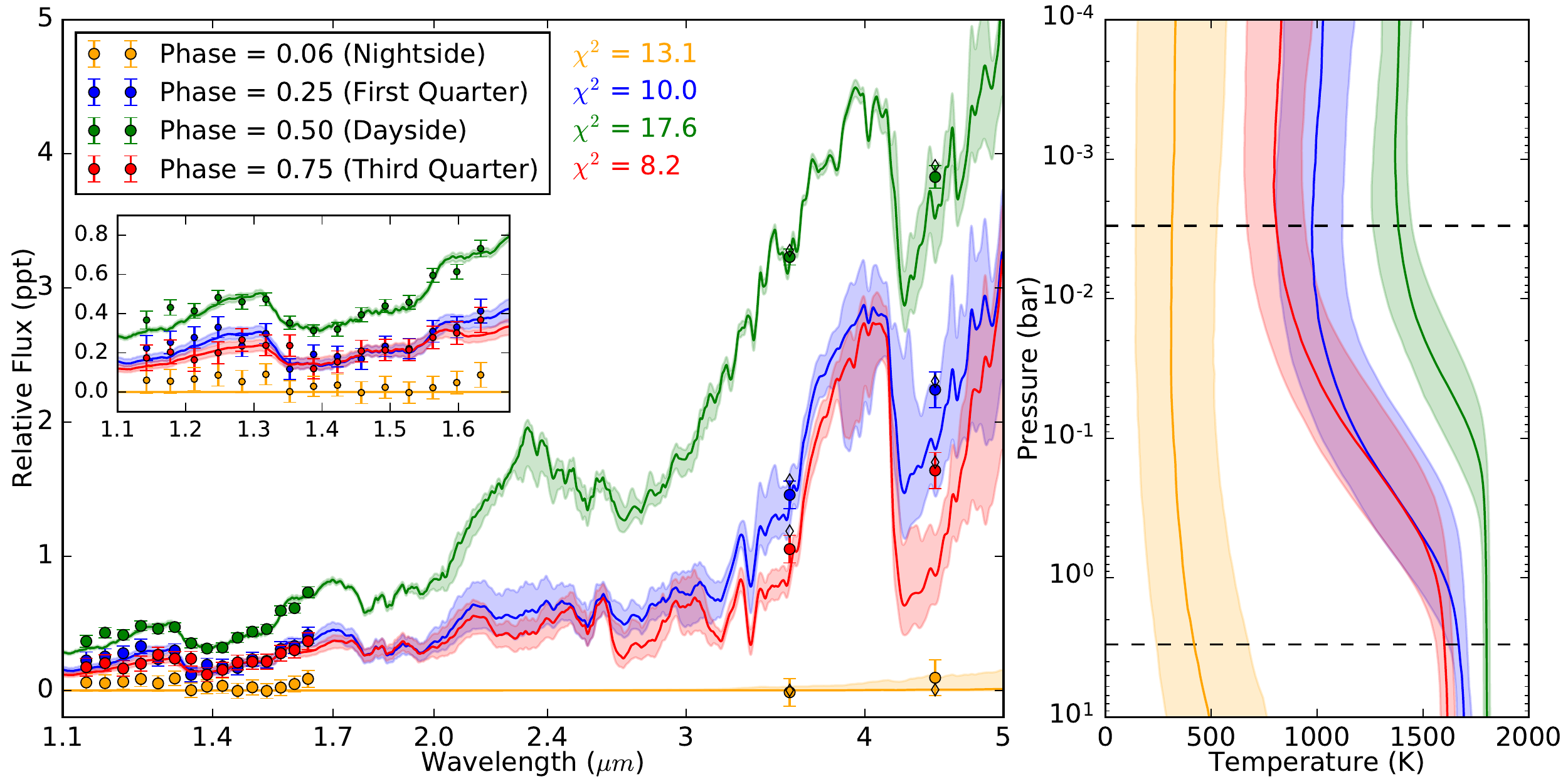}
\caption{Hubble Space Telescope WFC3 and Spitzer Space Telescope phase
  curve spectrum of WASP-43b at four different orbital phases (left) and the
  corresponding retrieved temperature-pressure profiles (right). 
Each set of colored circles depict measurements from the phase curve
observations, and the colored curves with shaded
regions represent the median models with 1$\sigma$ uncertainties.  The inset magnifies the WFC3 spectra.  
 H$_2$O absorption is readily apparent in the spectra
at 1.4 microns where a significant drop in the planet-to-star flux
ratio is apparent.  From \cite{2017AJ....153...68S}.  }
\label{fig:W43}       
\end{figure}

\subsection{Phase Curve Observations}
\label{sec:1.1.3}
In general, the most difficult atmospheric measurement to make with a
transiting exoplanet is that of the phase curve.  
The difficulty stems from the requirement to maintain high photometric
precisions of order 100 ppm (which, as with transits and eclipses, are
still necessary) over long timespans.  Typically, measurements last on
the order of the orbital period of the planet, which can be a day or
longer - compared to transit/eclipse events which last a few
hours. Furthermore, taking into account the very large telescope times needed, for
practical purposes phase curve observations thus far have been focused
on the shortest period planets with periods of 2 days or shorter.  

While early non-continuous observational attempts were made
\citep{2006Sci...314..623H},  a notable phase curve observation was made
by \cite{2007Natur.447..183K}, who observed HD~189733b with the Spitzer
Space Telescope at 8 microns during half an orbital period covering a
transit and eclipse.  The measured phase curve amplitude indicated a
modest day-night circulation, which results in more modest phase curve
amplitudes.  In addition, the brightest portion of the phase curve was
observed just before secondary eclipse.  For a tidally locked planet,
one may expect the hottest and thermally brightest point on the planet
to occur at the sub-stellar point.  However, strong atmospheric
winds (as predicted by \citealt{2002A&A...385..166S}) have the effect of
advecting the heat and hottest part of the planet westward of the
substellar point.   Longitudes westward of the sub-stellar point are
maximally viewable just before the secondary eclipse event itself,
leading to an observed light curve where the maximum flux is found
before an eclipse. 

An important aspect of phase curve observations done on transiting
planets is that a transmission spectrum and an eclipse spectrum can
also be derived from the same dataset, making it a particularly
constraining measurement.  An example can be seen in Figure
\ref{fig:W43} where a spectroscopic phase curve of WASP-43 was
observed by \cite{2014Sci...346..838S, 2017AJ....153...68S} with HST
and Spitzer.  In the phase-curve spectra, H$_2$O features were mapped
around the planet, and the emission spectrum showed strong absorption features.

\subsection{Accessible Transmission Spectra Exoplanets}
\label{sec:1.1.4}
In 2006, there were 158 known exoplanets, with the majority of
the population found from the radial velocity technique.  Moreover,
only a very small handful of exoplanets had their atmospheres
regularly detected, most notably HD~209458b and HD~189733b.  Not all
exoplanets are ideal atmospheric targets.  The more massive exoplanets
have smaller atmospheric scale heights due to the high surface
gravity, and will have smaller transmission spectral signals.  For
secondary eclipses, cooler exoplanets will have low thermal fluxes in
the optical and near-IR, so will be more challenging to detect at
those wavelengths.  

For an exoplanet transmission spectrum, a good indication of the
expected signal can be estimated by calculating the contrast in area
between the annular region of the atmosphere observed during transit
and that of the star.  The characteristic length scale of the
atmosphere is given by the pressure scale height,
\begin{equation}
H=\frac{k_B T}{\mu g}
\end{equation}
where $k_B$ is the Boltzmann constant, $T$ is the temperature of the
atmosphere, $\mu$ is the mean mass of atmospheric particles, and $g$ is the surface
gravity.  For giant exoplanets, the composition of the atmosphere can
be assumed to be dominated by a H/He mixture of near-solar
composition, which gives $\mu=2.3\times u$ where $u$ is the unified
atomic mass unit \citep{2008A&A...481L..83L}.  For transiting exoplanets, the
surface gravity is well known for a large majority of the prime
transmission spectral targets, as historically the radial velocity
method has been used to confirm the planetary nature of a transiting object.  Those exoplanet systems too faint to be
detected via radial velocity are
also often too faint to perform detailed atmospheric characterisation.  Moreover, knowing the mass of an exoplanet for transmission
spectroscopy is vital, as large degeneracies will be present if the
mass is unknown which will limit the usefulness in constraining the
atmospheric properties. 

\begin{figure}[t]
\sidecaption
  \centering
\includegraphics[scale=.53]{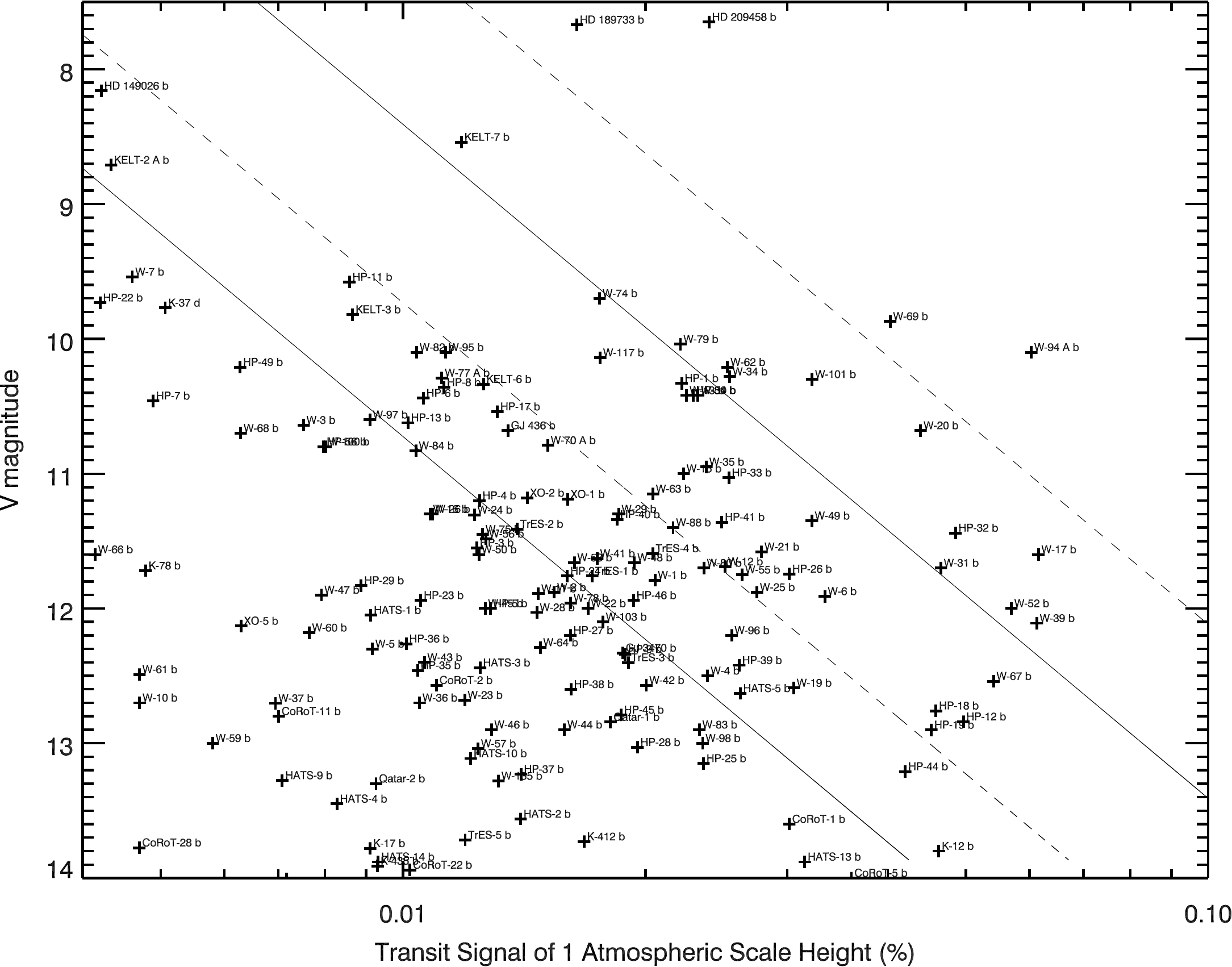}
\caption{Transmission spectral signal of 1 atmospheric scale height.
  Planets to the upper-right of the figure are easier to observe.  The
  lines indicate approximately constant S/N values. The exoplanet data was compiled
  from the exoplanets.org database.  The WASP and HAT-P planetary names have been abbreviated.} 
\label{fig:Transmission_targets}       
\end{figure}

 A good way to estimate the atmospheric temperature is to use the equilibrium temperature value, $T_{eq}$.
Assuming zero albedo and complete redistribution of heat around the planet, $T_{eq}$ can be
calculated using, 
\begin{equation}
T_{eq}=(1/4)^{1/4} T_{eff} \sqrt{\frac{R_{star}}{a}}
\end{equation}
where $a$ is the semi-major axis of the planet and $T_{eff}$  is the
stellar effective temperature \citep{2011ApJ...729...54C}.  The temperatures
derived thus far from transmission spectra have often been well within
these equilibrium values.  For example, HD~189733b, HAT-P-12b and
WASP-6b have temperatures derived from their transmission spectra of
1340$\pm$150 K, 1010$\pm$80 K and 973$\pm$144 K, respectively which compare
very favourably to their T$_{eq}$ values of 1200, 960, and 1150 K,
respectively \citep{2016Natur.529...59S}.  With the scale height
estimated, 
it is straightforward to approximate the absorption signal, $A$, of the annular area of one atmospheric scale height $H$ during transit, as
\begin{equation}
A = \frac{(R_{pl}+H)^2}{R_{star}^2}-\left(\frac{R_{pl}}{R_{star}}\right)^2,
\end{equation}
which can be further simplified assuming $H<<R_{pl}$ to,
\begin{equation}
A = \frac{2 R_{pl}H}{R_{star}^2}.
\end{equation}
Transmission spectral signals are typically on the order of 1 to $\sim$5$H$
in size, thus if the transit depth can be measured to about 1$H$ in
precision with sufficient spectral resolution, detectable spectral
features would begin to appear.  Plotting $A$ against the magnitude of
the host stars is a good proxy to compare the relative signal-to-noise
of different exoplanets.  While other factors such as cloud cover will
ultimately determine if atmospheric features will be present or not, all
other factors being equal, it is often a good guide to prioritise
exoplanets with the largest expected signal-to-noise values.  From
Figure \ref{fig:Transmission_targets}, the prominence of HD~189733b and HD~209458b become
apparent.  Both exoplanets orbit much brighter stars (V$\sim$7.7) than the
bulk of the known transiting planets, which dramatically improves 
the photon noise limits.  The ``puffiest'' planets, like WASP-17b may
orbit a much fainter star (V=11.6) but the expected atmospheric
transit signal is large ($A\sim$0.06\%) making it a comparable target
to HD~189733b in terms of expected S/N.  Perhaps 100 or more transiting exoplanets
are now accessible with today's instruments.  In practice, other practical considerations are
necessary to take into account.  For instance, ground-based
multi-object spectroscopy requires reference
stars to perform differential spectrophotometric
measurements.  However, few, if any suitable reference stars would
likely be available for observing a 7th magnitude target with a
typical 4 to 8 metre class telescope.  Such instruments have typical
fields of view of around 10 arc minutes, and it is unlikely that more
than one bright (and hence nearby) star would be close to each other
in the sky. 

\subsection{Exoplanets With Accessible Secondary Eclipses}
\label{sec:1.1.5}
\begin{figure}[t]
\sidecaption
  \centering
\includegraphics[scale=.53]{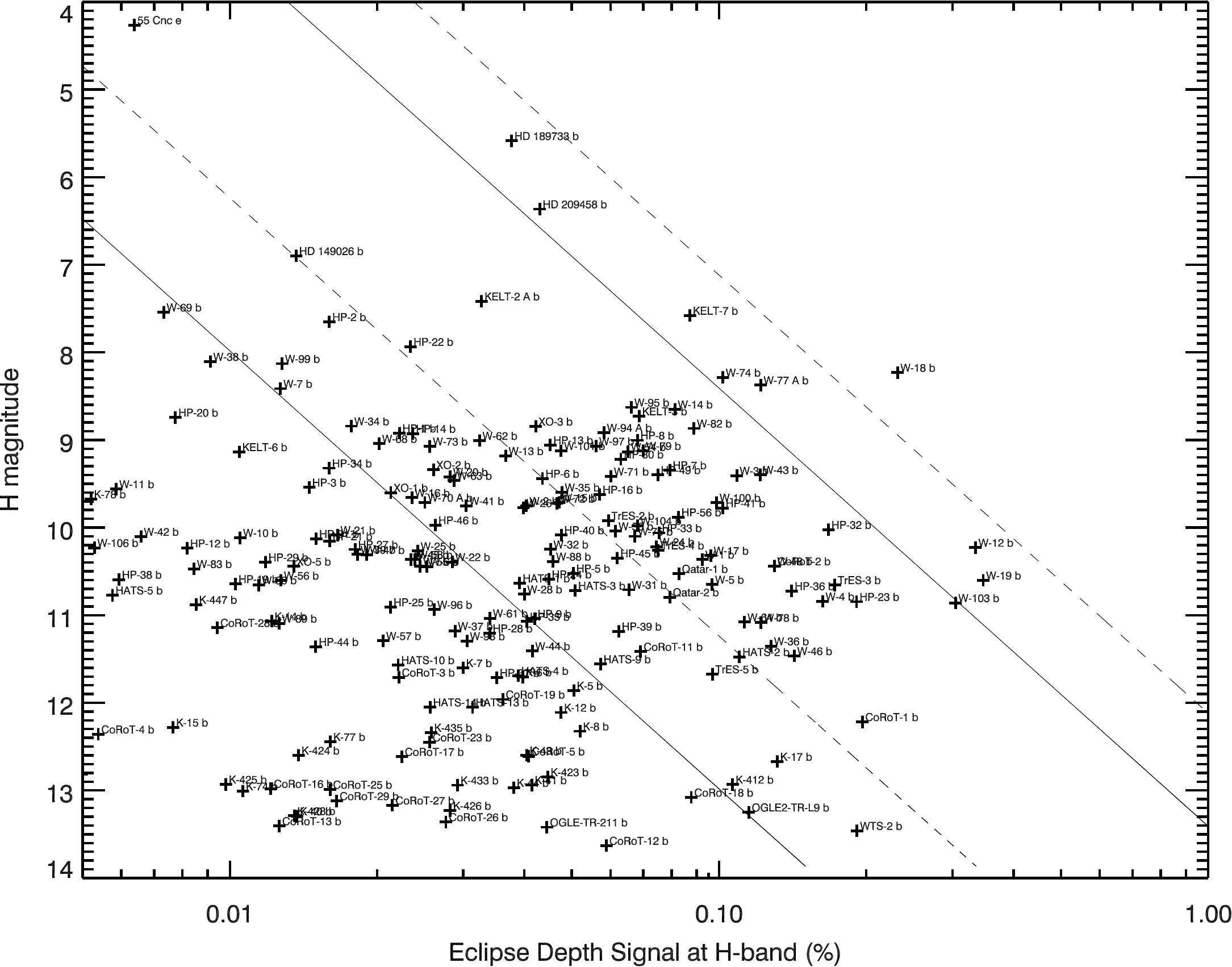}
\caption{Expected secondary eclipse signals at H-band.  Planets to the
  upper-right of the figure are easier to observe.  The lines indicate
  approximate constant S/N values. The exoplanet data was compiled
  from the exoplanets.org database.  The WASP and HAT-P names have been abbreviated.}
\label{fig:Eclipse_targets}       
\end{figure}

Not all exoplanets are favourable for secondary eclipse measurements.
The expected eclipse depths can be estimated in a similar exercise as
for transmission spectra. However, the results will be much more dependant
on the observed wavelength, because we measure the relative flux
contrast between the planet and the star, which radiate at very
different blackbody temperatures. 
The dayside
flux from the planet and
secondary eclipse depth, $\Delta f_{day}/f$ can be estimated from  
\begin{equation}
\frac{\Delta f_{day}}{f}=p_{\lambda}\left(\frac{R_{pl}}{a}\right)^2 + \frac{B_{\lambda}(T_{day})}{B_{\lambda}(T_{eff})}\times \left(\frac{R_{pl}}{R_{star}}\right)^2,
\end{equation}
which takes into account a reflection component with the wavelength
dependant albedo  $p_{\lambda}$, and a thermal component $B_{\lambda}$which is approximated here
assuming the planet radiates as black body with temperature $T_{day}$,
and the star also radiates as a black body with temperature $T_{eff}$
\citep{2010trex.book.....H}.  Similarly to Figure
 \ref{fig:Transmission_targets}, the expected secondary eclipse depth at a given wavelength can
then be plotted against the host star magnitude, at that wavelength
range, to assess the relative observability of different targets.  Of
course, in reality stars are not blackbodies, and neither are
planetary atmospheres so there can be significant deviations from
such simple estimations.  Nevertheless, such plots as shown in Figure
\ref{fig:Eclipse_targets}, can help illustrate the relative potential signal sizes between
planets, all else being equal.   

\subsection{Accessing the Atmospheres of Small Exoplanets}
\label{sec:1.1.7}

\begin{figure}[t]
\sidecaption
  \centering
\includegraphics[scale=.35]{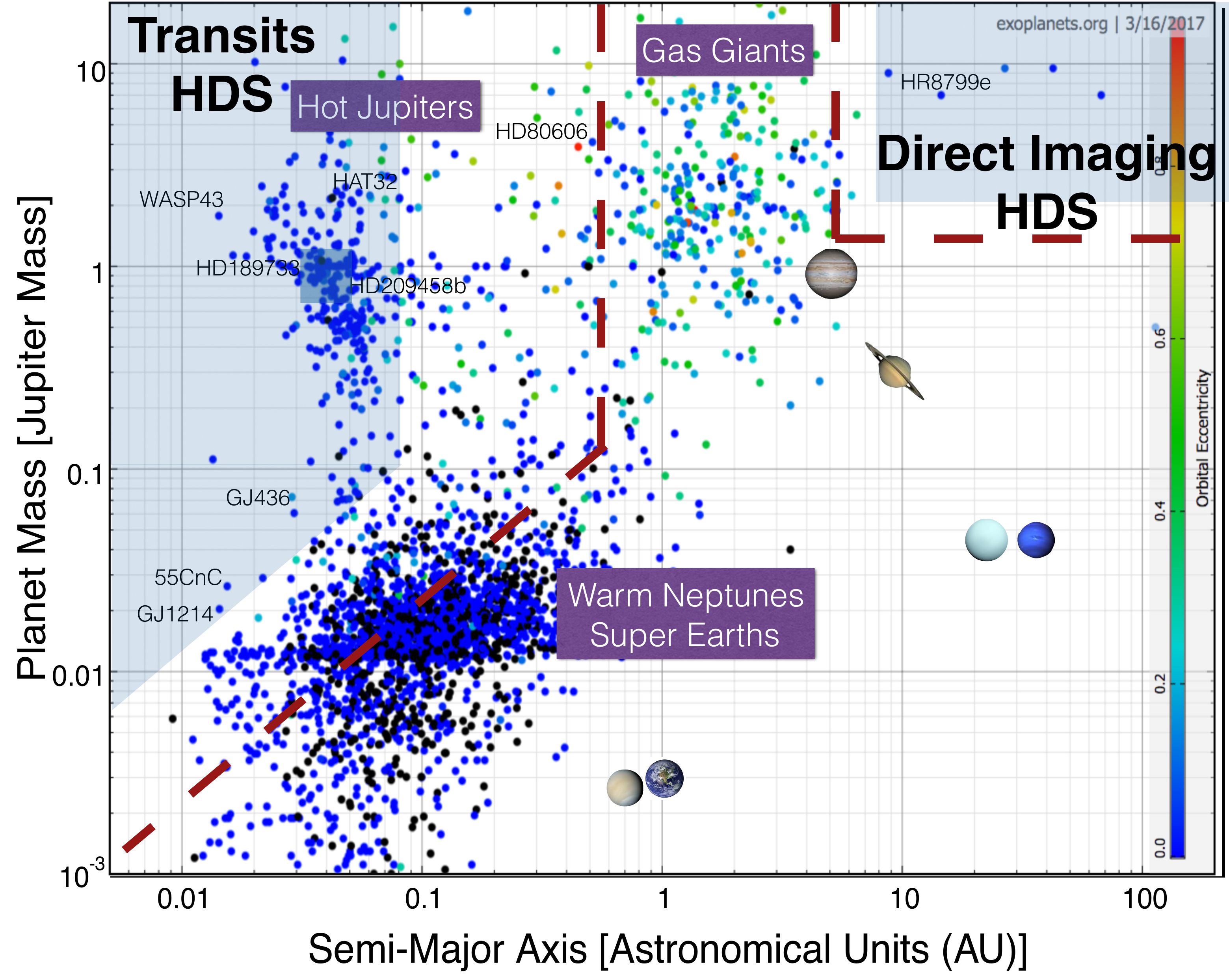}
\caption{Planet Mass vs. Semi-Major Axis for the detected exoplanets,
  compiled from the exoplanets.org database.  The broad exoplanet types
  are labeled and solar system planets also indicated.  The approximate current sensitivity of atmospheric
  studies is indicated (dark shaded region) and near-term expected
  improvements (red dashed lines).}
\label{fig:Mass_A}       
\end{figure}

Since the transit technique is predominately limited by the flux of
the host star, it can be used across a much broader wavelength range
than secondary eclipse measurements.  Transits have been measured in
the UV (e.g. \citealt{2003Natur.422..143V}) through to the far-infrared
(\citealt{2006ApJ...649.1043R}).  Additionally, much smaller exoplanets are
currently accessible with transit spectroscopy than with other
techniques.  There is strong interest to push characterisation down to smaller, cooler 
planets and so toward potentially life bearing worlds.  The task is
difficult, as small planets and cooler temperatures result in much
smaller transit atmospheric signatures (see Fig. \ref{fig:Mass_A}).  To overcome this difficulty,
currently there are two strategies being pursued to meet the near-term
goal of detecting atmospheric features and viably searching for
biomarker signatures within the atmospheres of extrasolar planets;
transit spectroscopy of exoplanets around small M-dwarfs (the 'M-dwarf
opportunity') or around very bright stars (the 'bright-star
opportunity').  These opportunities have driven dedicated transiting
M-dwarf searches, such as MEarth \citep{2009Natur.462..891C}, as well as dedicated space missions
such as TESS and PLATO which will search the brightest stars for
transits.

While searching for signatures of habitability will be an important
long-term exoplanet goal, it must be put into perspective as the very
wide and diverse group warm-Neptunes, super-Earths and small
terrestrial exoplanets represent a completely uncharted parameter
space. Developing a comprehensive theory 
to explain the atmospheres of these planets more generally, and
therefore put any atmospheric
detection within a wider context will represent an enormous challenge.  

The discovery of the super-Earth GJ1214b orbiting an M-dwarf
\citep{2009Natur.462..891C} provided the first atmospheric glimpses into
small exoplanets \citep{2010Natur.468..669B}.  
GJ1214b orbits a small, M4.5 dwarf star, which means it has a large
transit depth and atmospheric transmission signal.  Both of these
observable quantities scale inversely with the radius of the star
squared, so the signals of exoplanets orbiting smaller stars are
greatly enhanced. 
For GJ1214b, the star has a radius of $R_{star}$=0.2 $R_{\odot}$ making an
exoplanet signal $(1/0.2)^2$= 25$\times$ higher than if the same planet
orbited a sun-like star.  The
discovery of TRAPPIST-1b,c,d,e,f,g \citep{2017Natur.542..456G} has pushed the M-dwarf opportunity
to even smaller planets and stars, with the 0.114 $R_{\odot} $ sized M8 star
providing transmission spectral signals 77$\times$ higher than if the same
planet orbited a sun-like star.  To this end, even the atmosphere of
Trappist-1f, which is an Earth-sized exoplanet orbiting within the
habitable zone and is expected to be rocky, may potentially be
detected in the near-future.
Even with a large signal boost from a
small star, the atmospheric features are still expected to be small
and challenging to detect.  Current facilities such as HST have
allowed H-dominated atmospheres to be ruled out on several of the
TRAPPIST-1 planets \citep{2016Natur.537...69D}.

On the bright end, the discovery of a planet transiting around HD~97658b \citep{2013ApJ...772L...2D}
permitted atmospheric investigations of a super-Earth
around a much more massive K1V star.  In both the case of  HD~97658b
and GJ1214b, the exoplanet's atmospheres proved to be largely
consistent with heavy cloud-cover and no spectral features were
detected \citep{2014ApJ...794..155K, 2014Natur.505...69K}.

\section{Exoplanet Atmosphere Science Topics}
\label{sec:1.2}
In the following sections, I have highlighted a few current science
topics which transiting exoplanet atmosphere observations can (or will
hopefully) address.  While there are a very wide and diverse range of
science topics, many fall under a few basic categories which are
briefly described below.   

\subsection{Planet formation}
\label{sec:1.2.1}
Gas giant exoplanets are predominantly composed of a H/He mixture,
which was accreted during the planets formation from the
protoplanetary disk.  As such, the gas is primordial in nature so may
be expected to contain records of the formation conditions in which
the planet formed.  As transiting exoplanets are amenable to
spectroscopic studies, one may expect then to probe what would be
essentially the primordial gas and gain insights into the planet
formation process.

\begin{figure}[t]
  \centering
\includegraphics[scale=.75]{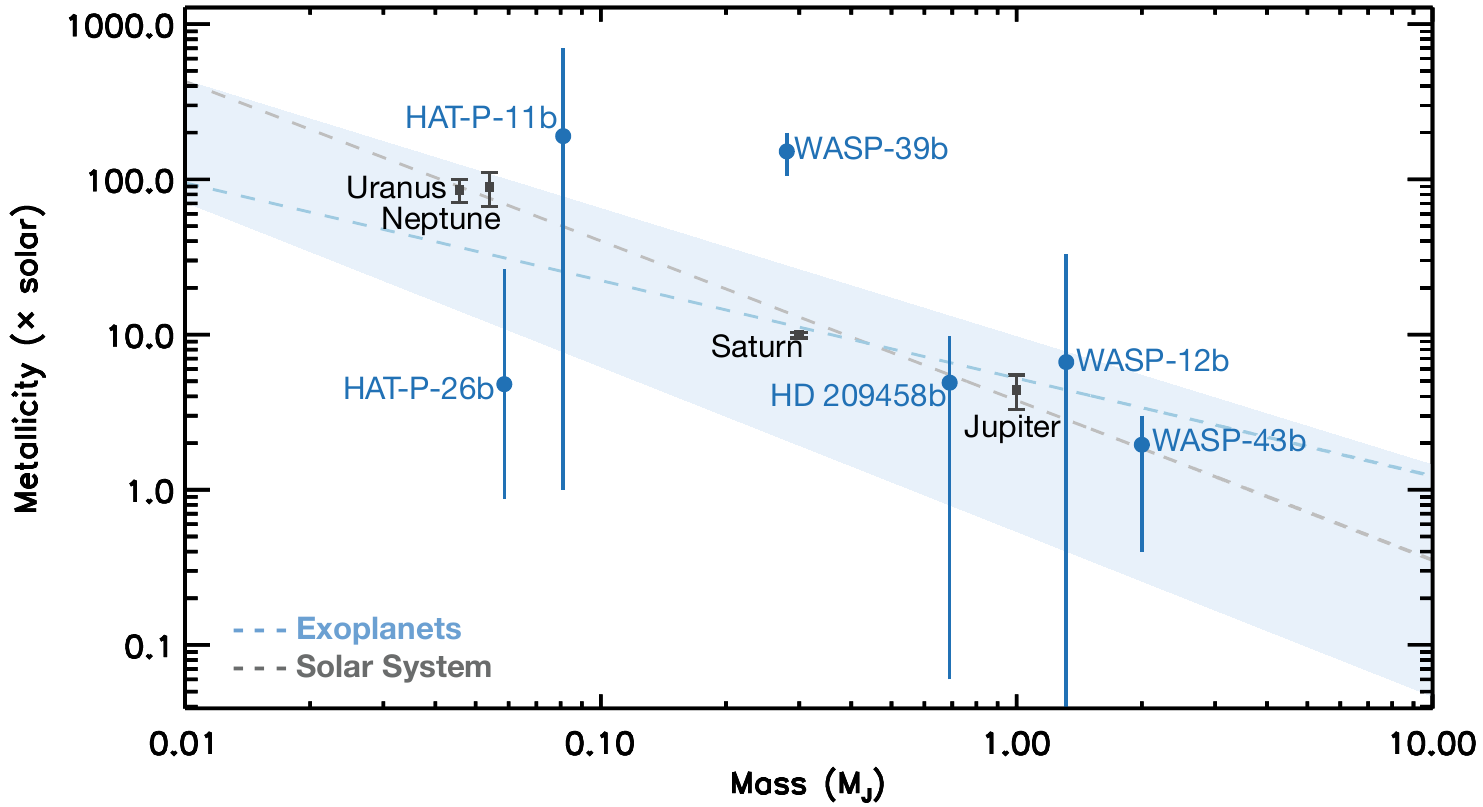}
\caption{Mass-Metallicity trend for solar system planets and
  exoplanets, adapted from \cite{2017Sci...356..628W} to include
  measurements of WASP-39b \citep{2018AJ....155...29W}, and HD~209458b \citep{2016AJ....152..203L}.}
\label{fig:Mass_metal}       
\end{figure}

There are two widely considered theories for how gas giant planets
form: gravitational instability and core accretion. Gravitational
instability is said to occur when the protoplanetary 
disk rapidly cools and collapses into planetary-mass fragments \citep{1997Sci...276.1836B}. Planets formed
via this mechanism would have the same bulk compositions as their local protoplanetary
disk material and their host stars. Alternatively, in the core-accretion model, giant planets
form in a multi-step process: first, sticky collisions of planetesimals lead to the formation of
protoplanetary cores; then, once the cores reach a threshold mass they accrete nearby gas in
a runaway fashion \citep{1996Icar..124...62P}.  Population synthesis
models from \cite{2012A&A...541A..97M}
and 
\cite{2013ApJ...775...80F}
suggest that in the core accretion paradigm, as the mass of
a planet decreases, its atmospheric metallicity increases. This is because lower mass planets
would be unable to accrete substantial gas envelopes, and thus would be more susceptible
to pollution by in falling, higher metallicity planetesimals. The giant solar system planets
agree with the latter scenario (see Fig. \ref{fig:Mass_metal}), as the metallicities derived from the methane abundance of
Jupiter (from the Galileo probe: \citealt{2004Icar..171..153W}),
Saturn, Neptune, and Uranus (from
infrared spectroscopy: \citealt{2009Icar..199..351F,2011Icar..211..780K, 2011Icar..215..292S}, respectively) show decreasing metal enhancement with increasing planet mass. Via
these two theories, exoplanetary atmospheres will exhibit different atmospheric properties
which can be measured from transmission and emission spectroscopy.  Gravitational instability theory
suggests that planets will have the same atmospheric metallicity as the central star, while
in core accretion theory lower mass planets will have higher
atmospheric metallicity.

Gas-giant exoplanets are widely expected to have retained the bulk of their primordial
atmospheres, and measuring the atmospheric abundances across a wide range of
planet masses should provide insight into formation mechanisms. \cite{2014ApJ...793L..27K}
found evidence the 2$M_{J}$ hot Jupiter WASP-43b follows
the same inverse mass-metallicity relationship as the solar system
planets. However, the Neptune-mass exoplanet 
HAT-P-26b has a measured water abundance at just 4$\times$ solar \citep{2017Sci...356..628W} which
is below the trend and suggests a different formation and/or
evolutionary processes. HAT-P-26b is consistent with recent envelope
accretion models \citep{2016ApJ...817...90L}, which argue 
that most hot Neptunes accrete their envelopes in situ shortly before their disks dissipate. In
both studies, the retrieved atmospheric water abundance was used as a proxy for the overall
planet's metallicity but the abundant carbon-bearing molecules will need to be measured
before stringent metallicity constraints are available.

The abundance ratio between carbon-bearing molecules and oxygen-bearing molecules
(C/O) is also expected to play a key role in constraining planet formation and migration
mechanisms \citep{2014ApJ...794L..12M}. The C/O ratio contains vital information such as the physical properties of
the accretion disk in which the planet formed, and a planets location within that disk.
This is especially pertinent for hot Jupiters, which migrated to short orbital periods though
may have initially formed beyond the snow line (e.g. \citealt{2011ApJ...743L..16O}). Carbon-rich
atmospheres would point to scenarios where hot Jupiters were initially located beyond the
snow line and accreted primarily carbon-rich gas, while O-rich atmospheres would point
instead to accretion of primarily oxygen-rich solid material \citep{2017ApJ...838L...9E}.

\subsection{Atmospheric Physics}
\label{sec:1.2.2}

Transiting exoplanets represent a new novel laboratory in which to
test our models of atmospheric physics.   The temperatures, gravities,
and chemical compositions occupy a very wide and new range of physical
conditions, making observations of these planets capable of giving new
broad physical insights into how planetary atmospheres operate.
Current theories of hot gaseous planets contain many open questions
about their atmospheric characteristics (temperature, clouds, energy
budget, atmospheric escape), their chemical abundances, and how they formed and evolved.  All these questions are intertwined, and
by observing, characterising, and comparing many exoplanets
across a broad parameter space, progress on answering some of
those questions can be made.  

Transiting planets orbit close to their host stars, making them 
tidally locked and highly irradiated. Those factors affect the
planet's vertical and horizontal (day-to-night) temperature
structure, and induce photochemical processes in their atmospheres,
which do not occur in the most related astrophysical objects, isolated
brown dwarfs (BD).  Therefore, hot
exoplanets are completely new objects with a set of physical
processes that are uniquely challenging to theoretically model.   

From the first observations of secondary eclipse spectra, the thermal
structure of hot Jupiters has been an active area of theoretical
investigations and observational efforts.  One of the first planets
charaterized through secondary eclipse measurements showed evidence for
a thermal inversion and hot stratospheric layer \citep{2008ApJ...673..526K}.
A hot stratosphere is caused by strong optical absorbers, which absorb
stellar radiation at altitudes higher than they thermally radiate
energy, which heats the upper atmosphere and causes a stratospheric
layer \citep{2003ApJ...594.1011H, 2007ApJ...668L.171B, 2008ApJ...678.1419F}.
On the Earth,
UV absorption by ozone creates a stratospheric layer and most solar
system planets including Jupiter and Saturn have stratospheres
\citep{1969ApJ...157..925G, 1974ApJ...193..481W, 1974ApJ...187L..41R}.

The presence or absence of a stratosphere is expected to change the
global energy budget and atmospheric circulation and dynamics of the planet, making
their presence and theoretical understanding an important aspect of
their overall atmospheric makeup.
In highly irradiated gas giant exoplanets that lack a strong optical absorber, the incident stellar
irradiation is absorbed deep in the atmosphere, near pressures of 1 bar (see \citealt{2008ApJ...678.1436B}).
This pressure is close to the near-IR emission photosphere, resulting in a monotonically
decreasing temperature profiles and a lack of a stratosphere.  At these pressures, the expected wind speeds
in a hot Jupiter will be able to efficiently redistribute heat around
the entire planet, leading to modest day/night temperature contrasts.
With a strong optical absorber at high altitudes,
the local gas is radiatively heated by the incident stellar flux,
creating a stratosphere. In addition, the winds at these lower pressures (higher altitudes) are not able to efficiently
redistribute the energy at their near-IR photospheres, creating a very strong day-night
temperature contrast. With a stratosphere, the hottest part of the planet becomes located at the highly irradiated
sub-stellar point, while the atmosphere becomes cooler towards the
limb. 

Several candidates for strong optical absorbers at altitude were
proposed \citep{2003ApJ...594.1011H, 2007ApJ...668L.171B, 2008ApJ...678.1419F} with
TiO/VO being the currently leading candidates.  \cite{2008ApJ...678.1419F} highlighted the importance of gaseous TiO and VO to
the optical opacities of highly irradiated hot-Jupiters, proposing two classes analogous to M
and L-type dwarfs. In this scenario, hot-Jupiters warm enough to still have gaseous TiO and
VO were dubbed “pM Class” planets. This class contains temperature inversions, and
appears “anomalously” bright in the mid-infrared at secondary eclipse, as the stellar incident
flux is absorbed high in the atmosphere and emitted as thermal flux at near-IR wavelengths.
Theoretical models predicting the transmission spectra of pM class of
planets would be dominated in the optical by TiO opacity \citep{2008ApJ...678.1419F}. The optical
transmission spectra of cooler pL Class planets (lacking TiO) are thought to be dominated by
neutral atomic Na and K absorption, and lack hot stratospheres.

\begin{figure}[t]
  \centering
\includegraphics[scale=.3]{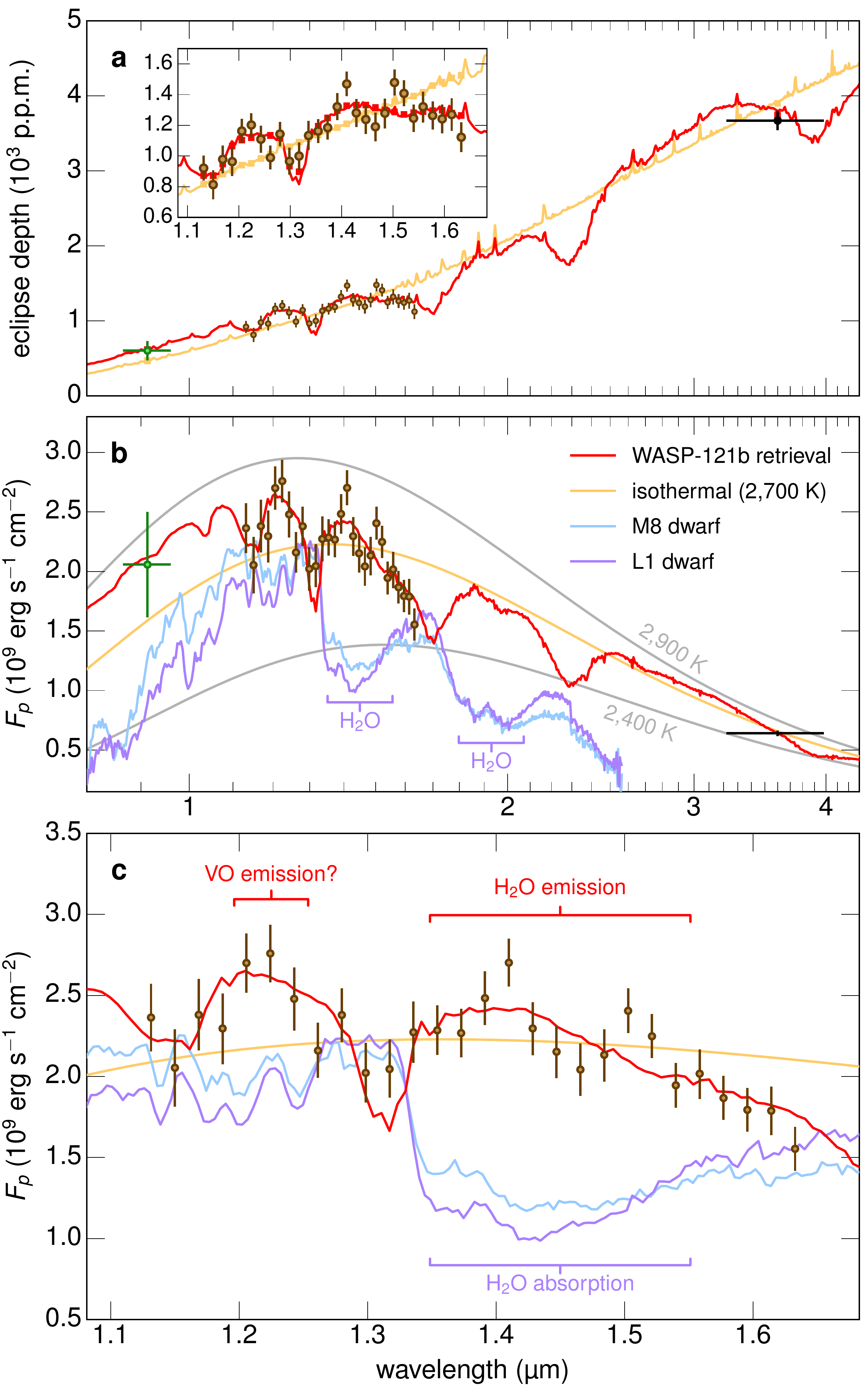}
\sidecaption
\caption{Emission spectrum of WASP-121b from \cite{2017Natur.548...58E}.
Shown is the measured HST, Spitzer, and ground-based eclipse
depths and the 1-$\sigma$ uncertainties, the horizontal error indicate
the photometric  bandpasses.  The yellow lines show the best-fit isothermal blackbody spectrum
with a temperature of 2,700 K, while the red lines show the best-fit
atmospheric model from a retrieval analysis.  The middle and lower
panels show isolated planetary flux with the stellar contribution
removed, and observed spectra for a M8
and L1 dwarf are shown for comparison (blue and purple lines), exhibiting
H$_2$O absorption bands. For
WASP-121b the H$_2$O band appears in emission. Spectra
for 2,400 K and 2,900 K blackbodies (grey lines) indicate the approximate
temperature range probed by the data. 
}
\label{fig:W121_inversion}       
\end{figure}

Follow-up studies of HD~209458b with more advanced data analysis
techniques did not support the presence of a stratosphere
\citep{2014ApJ...796...66D, 2015MNRAS.451..680E}.  In addition, despite
many dozens of exoplanets searched for signatures of stratospheres
with Spitzer, no definitive detections were made and confirmed.  For a
while it seemed TiO/VO may not be present in hot-Jupiter atmospheres.
\cite{2009ApJ...699.1487S} argued that vanadium oxide was not likely to fulfil this role due to
low abundances, and that the previously favoured titanium oxide would require unusually high
levels of macroscopic mixing to remain in the upper atmosphere.

A decade later, the topic of hot Jupiter stratospheres is still a hot
topic as a thermally inverted spectral signature was observed for
WASP-121b with H$_2$O seen in emission
(see Fig. \ref{fig:W121_inversion} and \citealt{2017Natur.548...58E}), and evidence for a stratosphere and
TiO seen in WASP-33b as well \citep{2015ApJ...806..146H, 2017arXiv171005276N}.  Compared to earlier studies, WASP-121b and WASP-33b are
much hotter (T$_{eq}>$2500 K) than the planets probed earlier, which
could indicate much hotter temperatures are required than earlier
theoretical studies indicated \citep{2008ApJ...678.1419F}.  However, it
remains unclear why some planets would have stratosphere layers, while
other seemingly similar very hot planets do not.

\subsection{Clouds and Hazes}
\label{sec:1.2.3}
Cloud and haze aerosols are ubiquitous for the planets with significant
atmospheres within our own solar system.  For hot exoplanets currently
amenable to transit characterization, clouds and hazes have also been found.
The atmospheric temperatures of the hot Jupiters are close to the condensation temperatures
of several abundant components, including silicates and iron.  The
formation of condensate clouds and
hazes is a natural outcome of chemistry in much the same way H, C, and O
combines to form H$_2$O, CO, and CH$_4$. The possible
presence of such condensation clouds was considered early on
\citep{2000ApJ...537..916S}.  Cloud and haze aerosols can form via
condensation chemistry, or alternatively the aerosols 
may be photochemical in nature (e.g. \citealt{2008A&A...485..547H, 2013cctp.book..367M}). Silicate and high-temperature cloud
condensates are expected to dominate the hotter atmospheres, while in cooler atmospheres sulphur-bearing
compounds are expected \citep{2010ApJ...716.1060V, 2012ApJ...756..172M, 2015A&A...573A.122W}.  The presence or
absence of clouds and hazes have strong implications on all aspects of a planet's atmosphere
including the radiation transport, chemistry, total energy budget, and advection \citep{2013cctp.book..367M}. As such, the presence of clouds and capacity to model them is currently a major uncertainty
and limitation in our ability to interpret exoplanet spectra and retrieve accurate molecular
abundances.

\begin{figure}[t]
  \centering
\includegraphics[scale=.65]{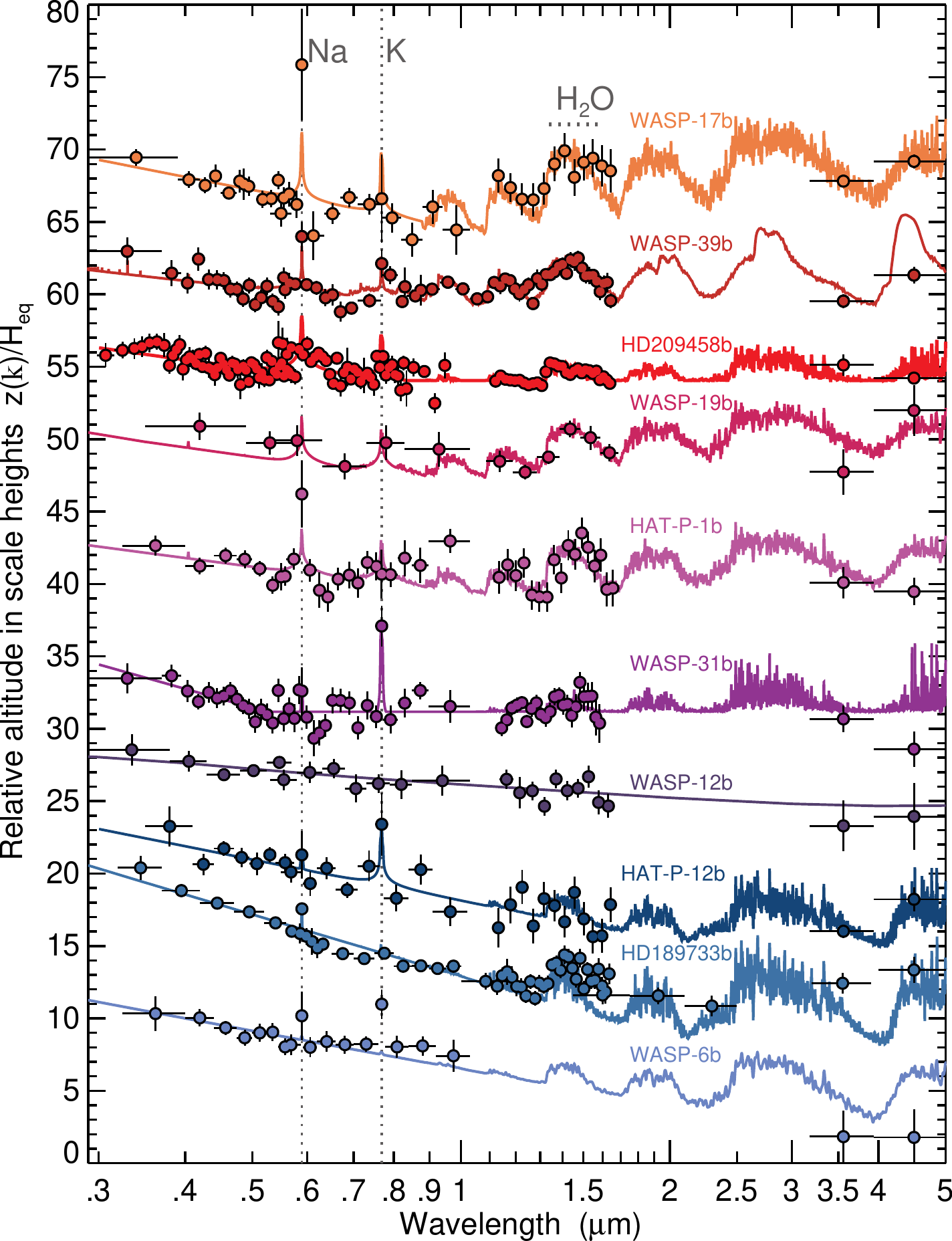}
\caption{Adapted from \cite{2016Natur.529...59S}.  Shown is the HST/Spitzer
  transmission spectral sequence of hot-Jupiter survey targets, with
  data from Wakeford et al. (2018) for WASP-39b also included.   Solid coloured lines show
  fitted atmospheric models and the prominent spectral features including
  Na, K and H$_2$O are indicated.  The horizontal and vertical error bars indicate
  the transmission spectra wavelength bin and 1$\sigma$ uncertainties,
  respectively.  Planets with predominantly clear atmospheres (towards
  top) show prominent alkali and H$_2$O absorption, with infrared radii values
  commensurate or higher than the optical altitudes.  Heavily hazy and
  cloudy planets (towards bottom) have strong optical scattering slopes,
  narrow alkali lines and H$_2$O absorption that is partially or
  completely obscured.}
\label{fig:HJSeq}       
\end{figure}

According to models, condensates would weaken spectral features, or mask some of
them, depending on the height of the cloud deck \citep{1999ApJ...513..879M,
2003ApJ...588.1121S, 
2005MNRAS.364..649F}.   In transmission spectra, a grey cloud (for example) can mask all absorption features below the
altitude of the cloud deck, and is an explanation for the muted water feature of HD 209458b
\citep{2013ApJ...774...95D} and the absence of features on the super-Earth
GJ1214b \citep{2014Natur.505...69K}.  The HST transmission spectrum of HD 189733b was found to
contain a high altitude scattering haze \citep{2008MNRAS.385..109P}, which has since been
confirmed by multiple follow-up HST measurements 
\citep{2011MNRAS.416.1443S, 2012MNRAS.422.2477H, 2012MNRAS.422..753G,
  2013MNRAS.432.2917P}.  For HD~189733b, the haze covers the entire optical regime, with a Rayleigh
scattering slope masking all but the strong Na I line cores and likely extends into the
near-IR, covering or muting the water absorption features (see Fig. \ref{fig:HJSeq}).  Most of
the exoplanets charaterized thus-far show some levels of clouds \citep{2016Natur.529...59S}, though the strong diversity of cloud and haze covers
found thus far indicates there will be a sizeable population hot Jupiters with largely clear atmospheres,
especially in the infrared where the scattering opacity of hazes and
clouds is likely to become greatly reduced. 

\section{Analysing Transmission Spectral Data}
\label{sec:1.3}
In the following sections, some of the basic data analysis procedures 
that have been developed to handle transit and eclipse
spectrophotometry at high precision are reviewed.  Included is a broad overview of the reduction steps,
time series fitting methods, statistical tools, and methods to handle
different noise sources.  When pushing transiting exoplanet spectroscopy
to very high (few ppm) levels, all potential sources of
noise tend to matter, and even what may appear to be very minor effects can become important
limiting factors.  

\subsection{Pre-observation steps}
\label{sec:1.3.1}

An important but often overlooked aspect of observational astronomy
and transiting exoplanet characterization is the steps one
has to make and plan for well in advance of working with any new
dataset.  Most all good science starts with an idea, and a science
question to investigate.  For transiting exoplanet science, even
fifteen years after the first atmospheric detection there are still
no dedicated instruments designed from the beginning to perform the
difficult task of obtaining the 10's or even 100's of ppm level photometric precisions across
hour-long timescales necessary.  This is in stark contrast to a
dedicated instrument such as HARPS, which has proven $\sim$m/s
radial velocity accuracies can be reliably achieved. 
As such, the planing and executing of transit/eclipse data still
requires special care.  Space-based data remains the gold standard,
as the data quality is much more homogenous and the levels of
precision demonstrated are much more reliably obtained.  Nevertheless,
a common mistake often made is to be overly optimistic or unrealistic
in the levels of precision that can be achieved, and the size of
potential atmospheric signatures.  The current history of transiting exoplanet
atmospheres has shown that more often than not, the atmospheric signal sizes
observed are smaller than predicted, and the noise levels achieved are
usually larger than photon limited (which exposure time calculators
assume).  Such was the case for Na on
HD~209458b \citep{2002ApJ...568..377C} which was about 3$\times$ smaller
than predicted, and most of the H$_2$O features seen by HST to date as well
have also been considerably smaller than predicted
(e.g. \citealt{2013ApJ...774...95D, 2013MNRAS.435.3481W, 2014Natur.505...69K}).  Such has the case been with secondary eclipse
measurements as well, with the \cite{2008Natur.456..767G} Spitzer
spectrum of HD~189733b showing H$_2$O features which were much smaller
than earlier models (e.g. \citealt{2008ApJ...678.1436B}).  Much of the smaller
features are likely due to natural explanations such as clouds and
hazes, which are still not robustly handled by most theoretical forward
models given the complexity.  For an observer, these aspects need to be
kept in mind when planing one's observations to ensure detections can
be made even if photon noise is not achieved and the signals sought
for are smaller than expected.  After all, the end goal is (or should be) to obtain
robust, impactful results published in reputable journals, not just
to collect data.\\

\noindent Basic Pre-Data Procedure Steps:
 \begin{enumerate}
\item{Have great idea to solve important science question, linking it
    to potential observations.}
\item{Check to see if the appropriate observations can be acquired, and
  with sufficient S/N.  Do not be overly optimistic or unrealistic in
  what can be achieved, if so, even if telescope time is awarded and
  good data is acquired, luck will be needed to obtain impactful results.}
 \item{Write a great telescope proposal(s).}
 \item{Submit the proposal \& convince a skeptical allocation committee. \\
        If rejected, revise proposal and resubmit next call if the case remains strong.}
 \item{Plan the observations very carefully.}                                       
  \begin{enumerate}
         \item{Being paranoid of mistakes is often a good thing.}
          \item{Execute the observations very carefully.}                                       
         \item{Do not underestimate the value of calibration frames.}
         \item{Make use of your collaborator's expertise.}
         \item{Maintain high photometric precisions over continuous
             multi-hour timescales.  Most telescope operators are not used to the methods
             to obtain high photometric precisions (e.g. no dithering due to the relative nature of the measurement)
           and care must be taken to ensure the observations are
           executed properly.}
\end{enumerate}
 \item{Download/collect images.}
\end{enumerate}

\subsection{Initial calibration overview}
\label{sec:1.3.2}
After obtaining a time-series dataset, the first data-reduction steps
are much the same as any other method.  However, before one begins
reducing data it is a very good idea to inspect each of the data-frames
in detail.  A time-series dataset may contain hundreds or perhaps
thousands of images.  With these large numbers, it can be easy to
overlook subtle issues which may not be obvious after aperture
photometry is performed or the spectral trace is extracted.  Stretching
an image and adjusting the image contrast to view the high and low
count level features can reveal possible issues such as bad pixels,
and making time-series movies of the data or blinking frames can be a
good method to reveal and get a feel for potential issues such as
positional drifts, cosmic rays, and detector ghosts.  Unexpected features may appear
as well, as very unlikely asteroid or satellite crossing events have
been noticed in HST time series data.  

\begin{figure}[t]
  \centering
\includegraphics[scale=.45]{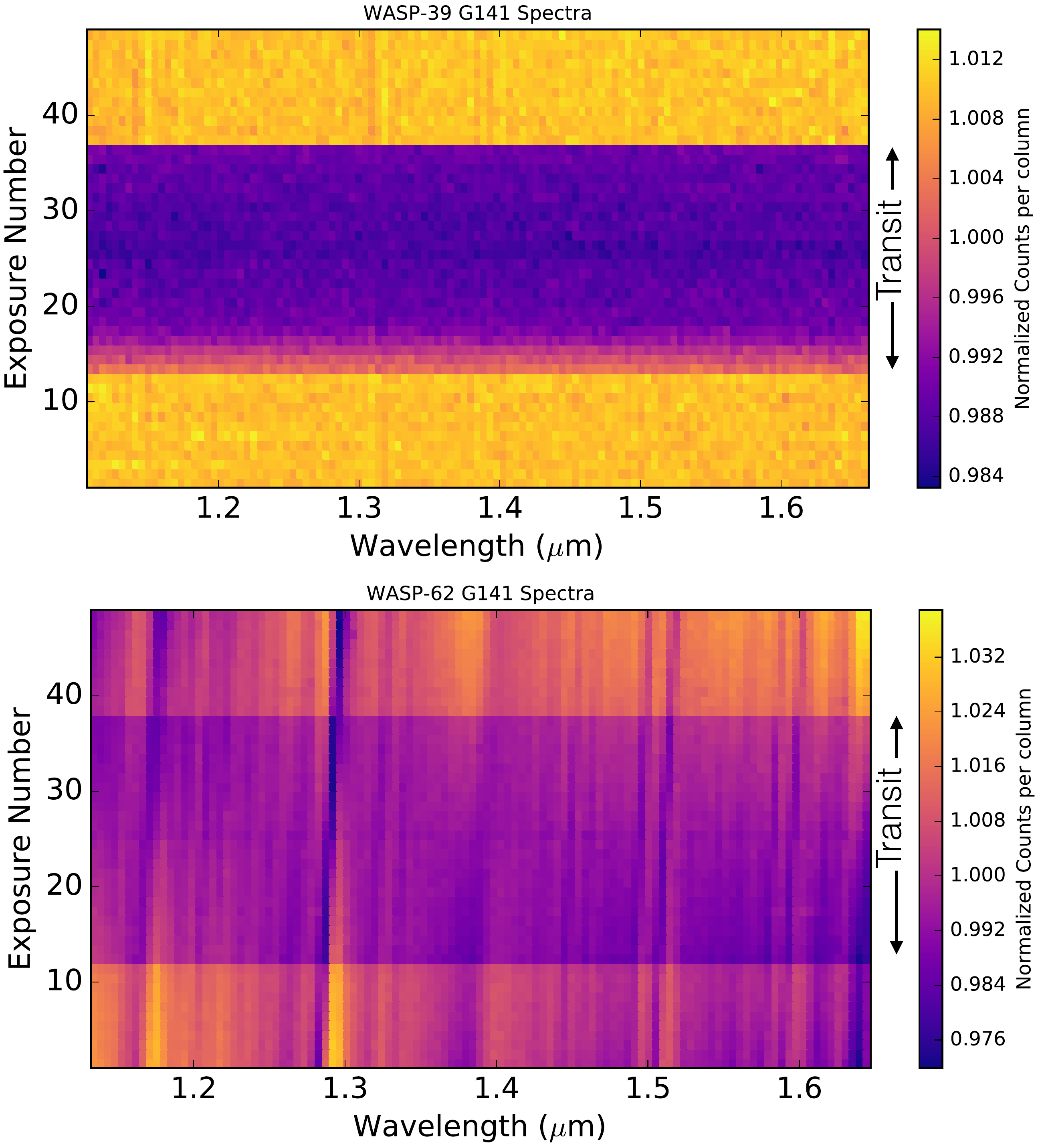}
\caption{Time-series spectral pixel maps for HST WFC3 G41 transit data of
  WASP-39b (top) and WASP-62b (bottom) used to visualise the overall data quality.  Plotted is the normalized count level at
  each pixel in the spectra vs time during the transit observation.  The transit event in both
  plots is evident by the drop in flux centred near the middle of both plots.  The WASP-62b transit suffered
from a guidestar problem which negatively impacted the telescope
guiding, and large drifts on the order of several pixels can be
seen. This problem negatively impacts the photometric quality of the
spectroscopic channels, as is apparent by the systematic trends which drift in
wavelength during the observation.  An example of a good quality dataset is
shown for WASP-39b (top) for comparison, which does not show such
wavelength-dependent trends.  Figs. courtesy of H. Wakeford.} 
\label{fig:pixelm}       
\end{figure}

For HST WFC3 data, time-series spectral pixel
maps can be utilised where a 2D histogram of the count
levels at each pixel in a spectra are plotted vs time (e.g. Fig \ref{fig:pixelm}).
As positional drifts are a major cause of systematic errors in WFC3
spatially scanned data, these maps help reveal the extent of such
drifts, and the presence of bad columns.

The first reduction steps with modern 2D image or spectral data are essentially the same as all other
traditional areas of astronomy.  For CCD data, the images are typically:
\begin{itemize}
\item{Trimmed of overscan regions, leaving the areas of the chip that
    contain useful data.}
\item{Individual bias frames are combined.}
\item{The flat-fields and science frames are processed to remove the overscan and average bias.}
\item{Bad pixel maps are constructed.}
\item{Flat-field images are combined and normalized.}\\
For time-series data, there is an important difference between
low-frequency features (large scale trends) in a flat-field, and
high-frequency pixel-to-pixel trends.  A widely adopted method for most all time-series
transit/eclipse data is to keep the
point-spread-function of the telescope on the same pixel (or
sub-pixel if possible) during the
entire course of the observations.  By doing so, low-frequency
flat-field features are not important and largely do not impact the
photometry.  A transit or eclipse light curve
is a differential measurement, and the absolute gain or count levels on
the detector are not utilised as each light curve is
individually normalized to the out-of-transit levels.  Thus, many
observers choose not to apply a flat-field correction, as it has been
sometimes been seen to introduce noise (e.g. \citealt{2017MNRAS.467.4591G}).
A successful flat-field correction, however, can correct for
high-frequency pixel-to-pixel variations.  These smaller scale gain
differences between neighbouring pixels can become important if the
telescope pointing is not entirely stable during the night, or if
there are significant seeing changes.  In these cases a flat-field
correction may prove important.  Given the extremely high count levels
of the science frames that are obtained during a transit, very large numbers
of well exposed flat-field images are needed such that the photon
noise levels per-pixel in the combined flat-field is comparable to the
integrated photon noise levels per pixel in the science frames.  Thus,
be sure to obtain as many flat-field and bias images as possible, as
it is unlikely taking one or two flat-field images would prove useful
when a time-series dataset is aiming for high photometric precisions
measuring a transit depth over perhaps hundreds or thousands of images.
\item{The science exposures are flat-fielded, and bad-pixels corrected
    if desired.}
\item{Cosmic rays are cleaned.}
\item{The spectra center is determined, trace defined, and 1D spectra
  extracted.\\
In a 2D spectral image, the center of the PSF at
  each pixel along the cross-dispersion direction is determined,
  called the trace. The background region is then defined and the counts in the cross-dispersion direction are then summed in an aperture of a
  given size along the trace to add up the total counts for each
  wavelength-pixel in the spectrum, typically subtracting the
  background.  The optimal aperture size and
  optimal background region have to be explored to
  optimise the time-series photometry at later stages.}
\item{A wavelength solution is determined.\\
Typically, arc lamp frames are
    gathered and the wavelength of specific emission lines are
    identified, which then provides a direct wavelength-to-pixel
    mapping.  In the case of transit spectroscopy, stellar absorption
    features can also provide a direct and unambiguous identification of the
    wavelength (for example the Na D lines in optical spectra).}
\end{itemize}

\begin{figure}[t]
  \centering
\includegraphics[scale=.4]{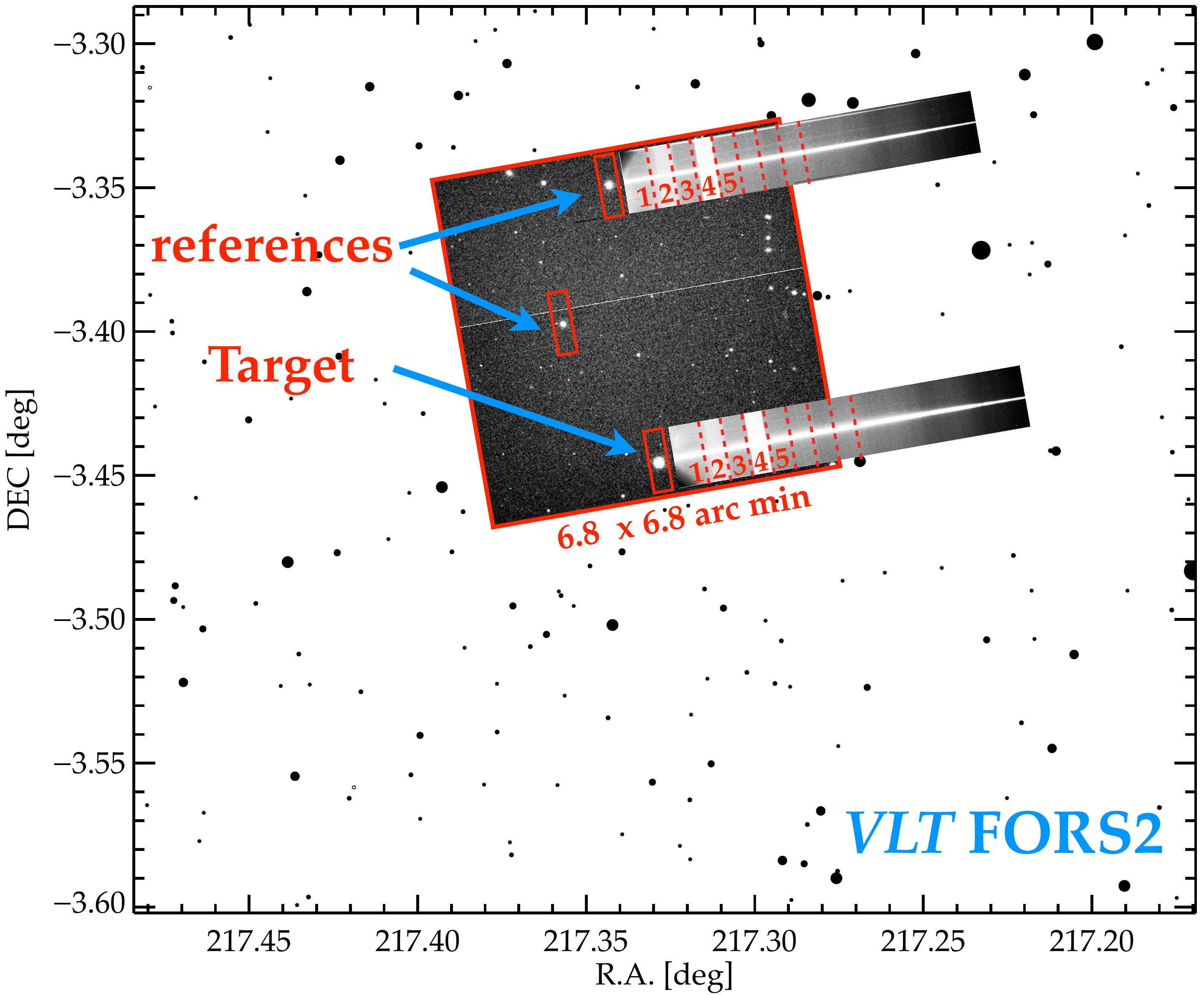}
\caption{Illustration of ground-based multi-object differential
  spectroscopy.  Shown is the field of view of VLT FORS2 instrument
  along with the surrounding field of WASP-6b, overlaid with an
  acquisition image and a
  spectroscopic image.  The red rectangles on the acquisition image show the large slits as they are
  projected on the sky, which encompass the target and a number of
  nearby reference stars. The numbered rectangles depict example
  wavelength bins for the target and reference star.  Fig. courtesy of N. Nikolov.} 
\label{fig:W6fov}       
\end{figure}

\begin{figure}[t]
  \centering
\includegraphics[scale=.60]{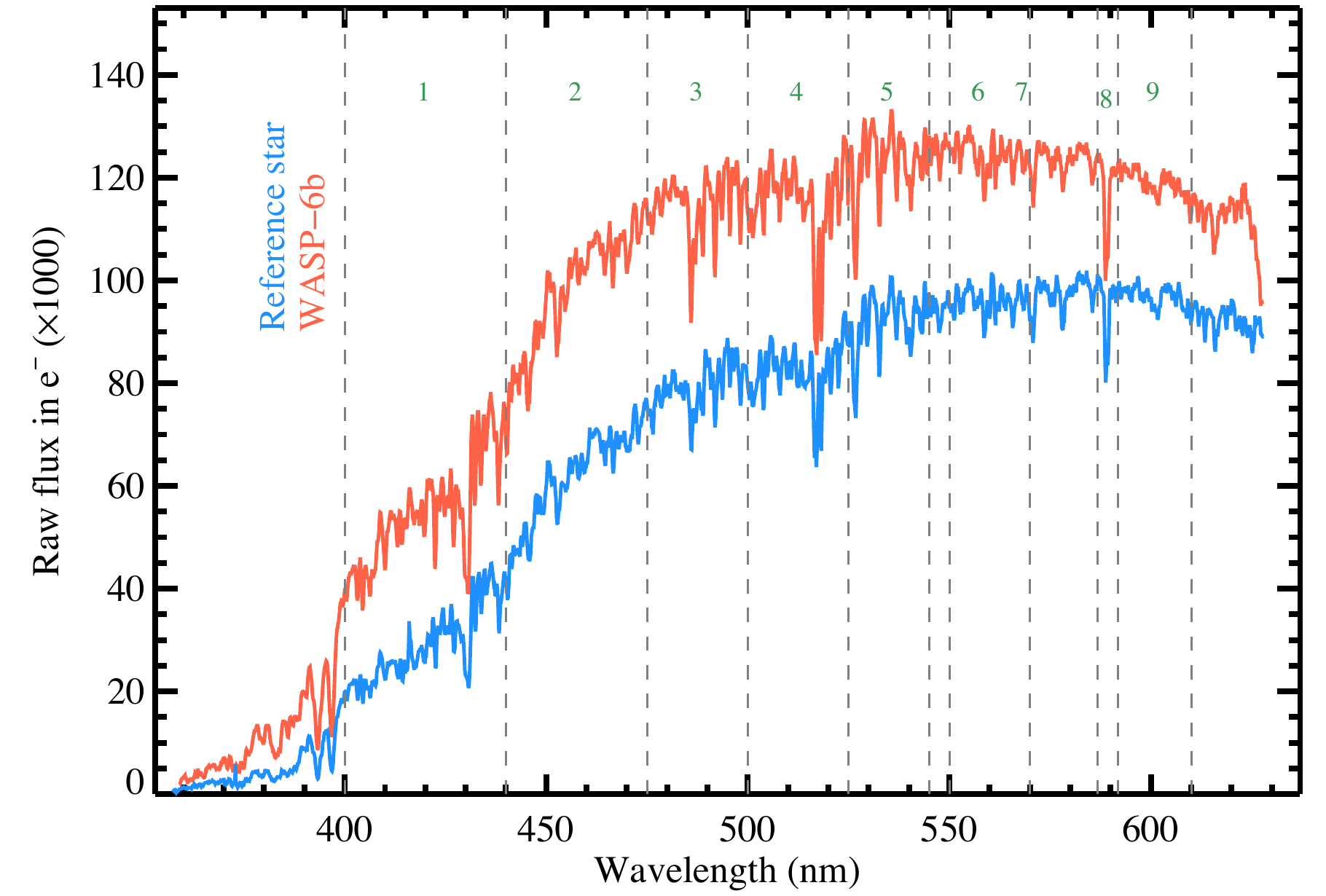}
\caption{Stellar spectra of WASP-6A and a target star, along with the
  bins used to create the spectroscopic channels transit light
  curves.  Fig. courtesy of N. Nikolov.} 
\label{fig:W6star}       
\end{figure}

With space-based transit spectroscopy, one can proceed more or less directly from
spectral extraction to fitting light curves.
In the case of ground-based multi-object spectroscopy (MOS), the spectra of
two or more stars must be extracted as a reference star is needed to
correct each spectroscopic channel for the effects of Earth's
atmosphere.  The method of MOS is essentially an extension of differential
photometry, and was initially applied by \cite{2010Natur.468..669B} to observe the
transmission spectra of GJ1214b with VLT FORS2.  In MOS observations,
the spectra of two or more stars are obtained using typically a mask
or longslit, with the slit sizes specifically chosen to be very wide
such that slit light losses are minimized or eliminated all together
(see Fig. \ref{fig:W6fov}).
Sizes of 10''+ or more are often used, with an example of \cite{2016ApJ...832..191N}
using 22''$\times$90" sizes slits on VLT FORS2.  The target and
reference spectra must then be accurately wavelength calibrated, such
that light curves at the same wavelengths can be binned and constructed for use
in differential photometry (see Fig. \ref{fig:W6star}).  In principle, the reference star should
have all the adverse effects from Earth's atmosphere also encoded in
the light curves including seeing variations, transparency
variations, and changing atmospheric extinction.  Dividing the target
star light curve by the reference star then largely subtracts out
these features.  In practice, other effects must also be dealt with
such as instrument flexure and pointing drifts.

\subsection{Pre-light curve fitting}
\label{sec:1.3.3}

For most areas of astronomical observations, the data reduction steps
would largely stop here and the scientific analysis would begin, though with transmission or emission
spectroscopy of exoplanets, the job is just partially complete.  The very
important steps of fitting the transit light curves (along with
systematic trends) for each
spectroscopic channel still need to be performed.  In most cases, the
light curve fitting stage takes the most time and effort and it
remains a non-trivial process which must be handled with care.

There are several important initial steps which need to be taken before
fitting transit light curves:\\
\begin{itemize}
\item{Calculate accurate time stamps for the time series images.}\\
 Very high quality light curves can often have quantities like the central transit time
measured to precisions of seconds or better, and other studies will
often use the measured quantities to look for transit time variations
or followup an exoplanet at a later date, making accurate time stamps
very important.  Indeed, care must be taken when combining different datasets from different
observatories, as it is often hard to know exactly what clock standard
was used.  The commonly used Julian Date (JD) can be specified
in several different time standards which are often unreported, complicating
timing studies.  Fortunately, \cite{2010PASP..122..935E} provides a 
useful tool to calculate the Barycentric Julian Date in the
Barycentric Dynamical Time (BJD$_{\rm TDB}$) standard, which has
become a common practice in the exoplanet literature (http://astroutils.astronomy.ohio-state.edu).

With an accurate time-stamp, the projected separation between the
planet and star, $s$, as a function of the orbital phase, $\phi$, can be calculated as,
\begin{equation}
s(\phi)=\frac{a}{R_{star}}\sqrt{[\sin{(2\pi \phi})]^2+[\cos(i)\cos(2\pi \phi)]^2}
\end{equation}
where $a$ is the semi-major axis and $i$ is the inclination of the
orbit.  $s(\phi)$ is a quantity provided as input to widely used
transit models such as \cite{2002ApJ...580L.171M}.\\

\item{Align the spectra onto a common wavelength/pixel grid.}\\
Spectra are often seen to exhibit non-trivial shifts on the detector
during the time series sequence.  These shifts 
need to be accounted for if the spectra are to be placed
on a common wavelength grid such
that spectral bins can accurately be extracted.  Without applying such a
correction, fixed pixel bins of the spectral time series could contain a mix of
wavelengths as the light from neighbouring pixels would contaminate the
bin and degrade the light curves.  Shifts between spectra can be
easily measured using cross correlation procedures, and the spectra can
be interpolated onto a common scale.\\
\item{Calculate limb darkening coefficient for the wavelength bin of interest.}\\
Limb darkening is an essential component determining the shape of a
transit light curve, 
enhancing the U-shape of transit light curves due to the non-uniform flux profile across stellar discs.
Thus, an accurate treatment of stellar
limb darkening is critical when deriving precise planetary radii and
measuring transmission spectra.  The effects of limb darkening are
typically parameterized using a specified functional form (or law),
and theoretical limb-darkening coefficients (LDCs) are calculated using stellar models (e.g. \citealt{2000A&A...363.1081C}). 
The most commonly used in exoplanetary transit work are:\\
the linear law
\begin{equation}  \frac{I(\mu)}{I(1)}=1 - u(1 - \mu),\end{equation}
the quadratic law
\begin{equation}  \frac{I(\mu)}{I(1)}=1 - a(1 - \mu) - b(1 - \mu)^2 ,\end{equation}
the three-parameter non-linear law, 
\begin{equation}  \frac{I(\mu)}{I(1)}=1 - c_2(1 - \mu) - c_3(1 - \mu^{3/2}) - c_4(1 - \mu^{2}), \end{equation}
and the four-parameter non-linear law
\begin{equation}  \frac{I(\mu)}{I(1)}=1 - c_1(1 - \mu^{1/2}) - c_2(1 - \mu) - c_3(1 - \mu^{3/2}) - c_4(1 - \mu^{2}), \end{equation}
where $I(1)$ is the intensity at the centre of the stellar disk,
$\mu=cos(\theta)$ (where $\theta$ is the angle between the line of sight and the
emergent intensity), while $u$, $a$, $b$, and $c_n$ are the LDCs.  These laws can all
be used along with the analytical transit light models of
\cite{2002ApJ...580L.171M} or \cite{2015PASP..127.1161K}.

\begin{figure}[t]
  \centering
\includegraphics[scale=.40]{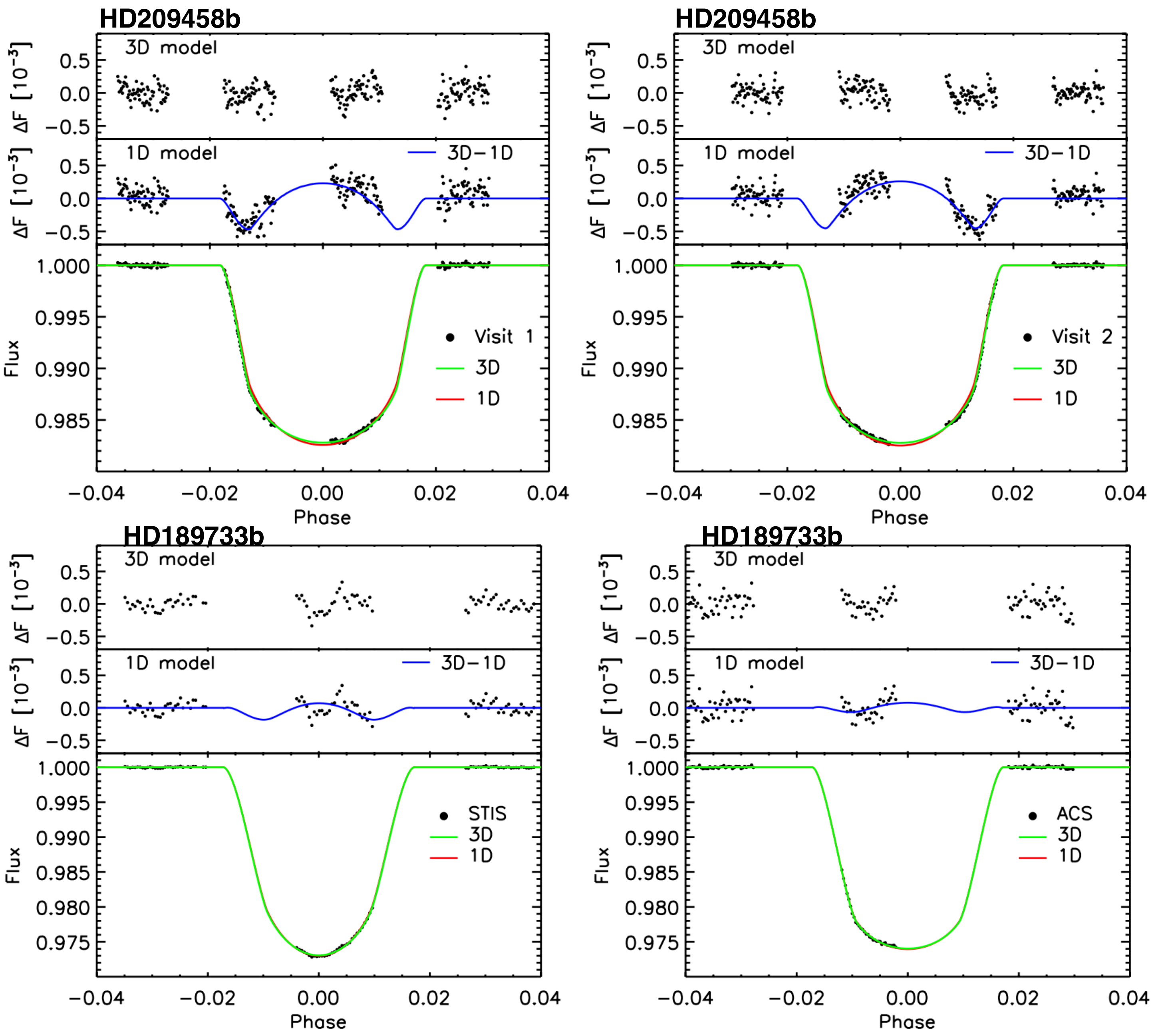}
\caption{HST transit light curves of HD~209458b and HD~189733b
  compared to transit model fits using 1D and 3D models (adapted from
  \citealt{2012A&A...539A.102H}).  (Top) For HD~209458b 1D models are unable to fully reproduce the
  transit shape, leading to a characteristic ``w'' shaped residual as
  seen in the middle panels for each fit and 3D stellar models do a better
job of fitting the transit.  (Bottom) The 1D models perform better for
HD~189733b.} 
\label{fig:LD}       
\end{figure}

Fitting for LDC from the transit light curves is widely used, with the
quadratic law most often adopted.  However, there are degeneracies in
fitting for the coefficients and without a proper treatment unphysical
stellar intensities can result which can bias the results.  This can happen especially for
grazing transits as the full stellar disk (and its intensity profile) is not sampled during the transit.
In addition, simple limb darkening laws can also do a poor job of reproducing a real stellar intensity profile.
For many transmission spectral applications, theoretical stellar
models 
have proven adequate for many transit light curve fits
(e.g. \citealt{2011MNRAS.416.1443S}) and the latest 3D models (see
Fig. \ref{fig:LD} and \citealt{2012A&A...539A.102H, 2015A&A...573A..90M}) have 
improved upon many of the deficiencies seen in earlier 1D models \citep{2007ApJ...655..564K, 2008ApJ...686..658S}.  For transit spectroscopy, it is often recommended to fix
the limb-darkening coefficients to their theoretical values, and
inspect the fitted residuals to see how well the stellar models are
performing, and fit for the coefficients if necessary.

\end{itemize}

\subsection{Light curve fitting}
\label{sec:1.3.4}

When fitting for a spectroscopic transit or eclipse dataset, the first
fits one typically performs is on the wavelength-integrated flux of the spectrum,
which is called the white-light curve.  The white-light curve fit
helps provide the overall system parameters such as $i$, $a/R_{star}$
and center of transit time $T_0$ as well as the average transit depth
across the wavelength range of the spectrum.   

In most all transit light curves to date, a model of the systematic
trends (any non-transit/eclipse related phenomena which affects the
light curve) must also be taken into account, whether they are of an 
instrumental or astrophysical nature.  For example, in Spitzer IRAC transit
photometry, it has been widely established the intra-pixel sensitivities
and pointing jitter cause variations in the photmetric light curves
which must be modelled and removed \citep{2006ApJ...653.1454M, 2008ApJ...673..526K}.  For HST STIS data, thermal breathing
trends cause the point-spread-function (PSF) 
to change repeatedly for each 90 minute spacecraft orbit around the Earth, producing corresponding photometric changes in the light curve
which results
in photometric changes in the light curve \citep{2001ApJ...552..699B}.  Systematic errors are
often removed by a parameterized deterministic model, where the
non-transit photometric trends are found to correlate with a number
$n$ of external parameters (or optical state parameters, $\mathbf{x}$).
These parameters describe changes in the instrument or other external
factors as a function of time during the observations, and are fit
with a coefficient for each optical state parameter, $p_n$, to model and remove (or detrend) the photometric
light curves.

When including systematic trends, the total parameterized 
model of the flux measurements over time, $f(t)$, can be 
modelled as a combination of the theoretical transit model, $T(t, \mathbf{\theta})$
(which depends upon the transit parameters $\mathbf{\theta}$), the total
baseline flux detected from the star, $F_0$, and the systematics error
model $S(\mathbf{x})$ giving,
\begin{equation} 
f(t)=T(t, \theta)\times F_0 \times S(\mathbf{x}).
\end{equation} 
In the case of HST STIS data, external detrending parameters including the 96
minute HST orbital phase, $\phi_{HST}$, the $X_{psf}$ and $Y_{psf}$ detector
position of the PSF, and the wavelength shift $S_{\lambda}$ of the
spectra have been identified as optical state parameters (\citealt{2011MNRAS.416.1443S}).  The optical
state parameters must be properly normalized such that they do not
contribute in changing the overall average system flux, and in the
case of STIS data a fourth-order polynomial with $\phi_{HST}$ has
been shown to sufficiently correct the instrument systematics such
that $f(t)$ can be written as,
\begin{multline} 
f(t)=T(t, \theta)\times F_0 \times
(p_1\phi_{HST}+p_2\phi_{HST}^2+p_3\phi_{HST}^3+p_4\phi_{HST}^4+1)\\
\times(p_5S_{\lambda}+1)\times(p_6X_{psf}+1) \times(p_7Y_{psf}+1).
\label{eq:sys}
\end{multline} 
Note that the systematics model can also be additive rather than
multiplicative as in Eq. \ref{eq:sys}.
Determining what systematic model to use, identifying parameters
which successfully detrend light curves, and finding suitable functions while avoiding overfitting remains
a large non-trivial challenge when analysing transit light curves.  This is
especially true if the data collected does not have a significant
history and standard practices determined when
dealing with high precision photometric time-series measurements.
Non-parametric methods have also been developed to model
systematic models, and methods such as Gaussian processes (GP) have
the benefit of not imposing a specific functional dependence on the
optical state vectors (see \citealt{2012MNRAS.419.2683G}), and have become widely
used.

\begin{figure}[t]
  \centering
\includegraphics[scale=.34]{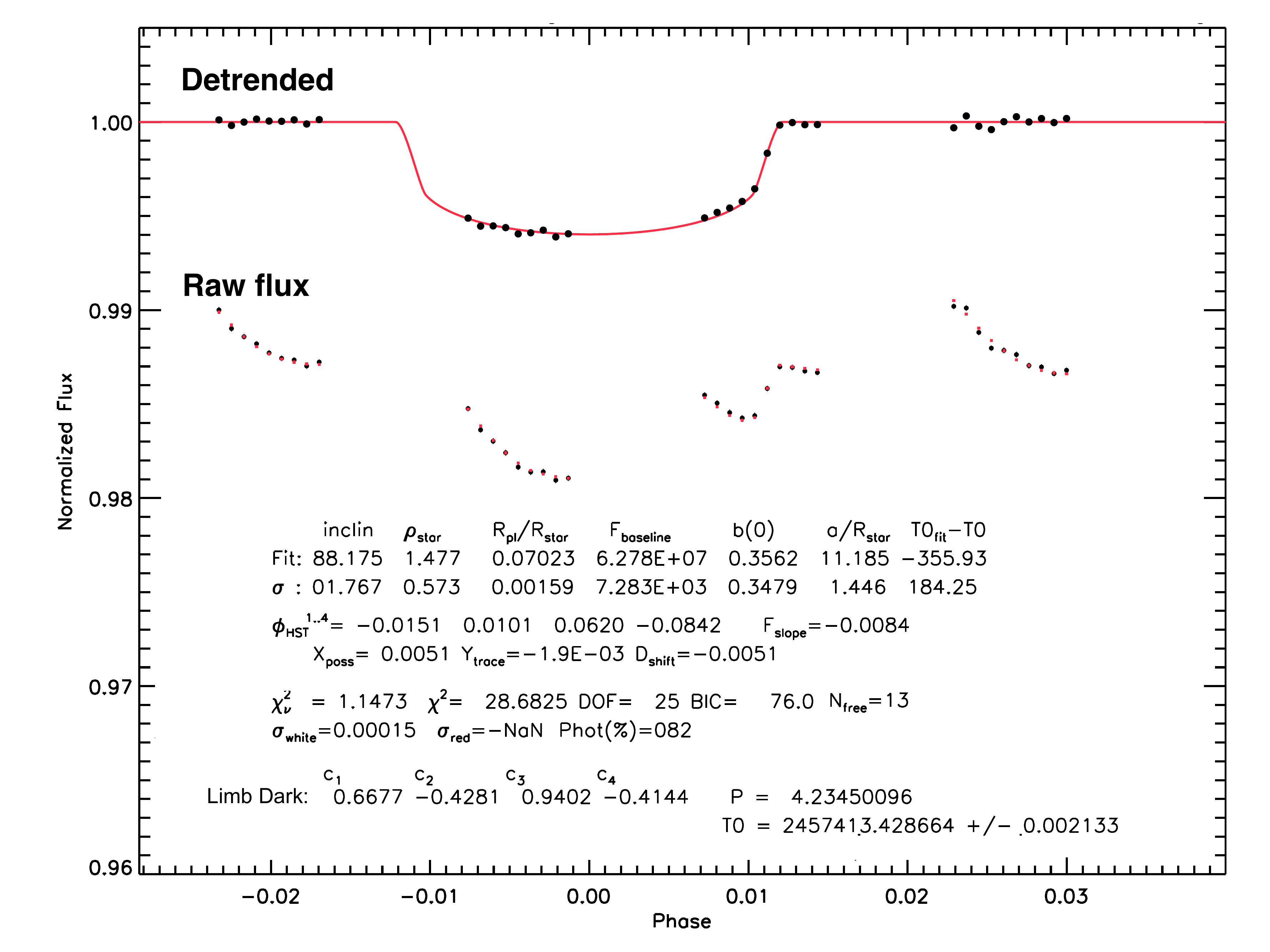}
\caption{Example transit light curve minimum $\chi^2$ fit to HST STIS
  data of HAT-P-26b (Wakeford et al. 2017).  Top shows
  the detrended normalized light data (black points) with the transit
  light curve model (red line),
  middle shows the raw flux which includes the instrument systematic
  trends along with the best-fit model (red points), an arbitrary offset has
  been applied for clarity.  The best-fit model parameters and several
  statistics of the model are also indicated. In this example, the fit
achieves precisions of 82\% the theoretical photon noise levels.}
\label{fig:chi2min}       
\end{figure}

When fitting transit models to the data, typically one begins by
finding a minimum $\chi^2$ solution (see Fig. \ref{fig:chi2min}) from routines such as
the Levenberg-Marquardt (L-M) least squares method \citep{2009ASPC..411..251M},
while Markov chain Monte Carlo techniques (MCMC) methods are useful
for deriving robust error estimates that can account for complicated
degeneracies between model parameters \citep{2013PASP..125...83E, 2013PASP..125..306F}.

\begin{figure}[t]
  \centering
\includegraphics[scale=.51]{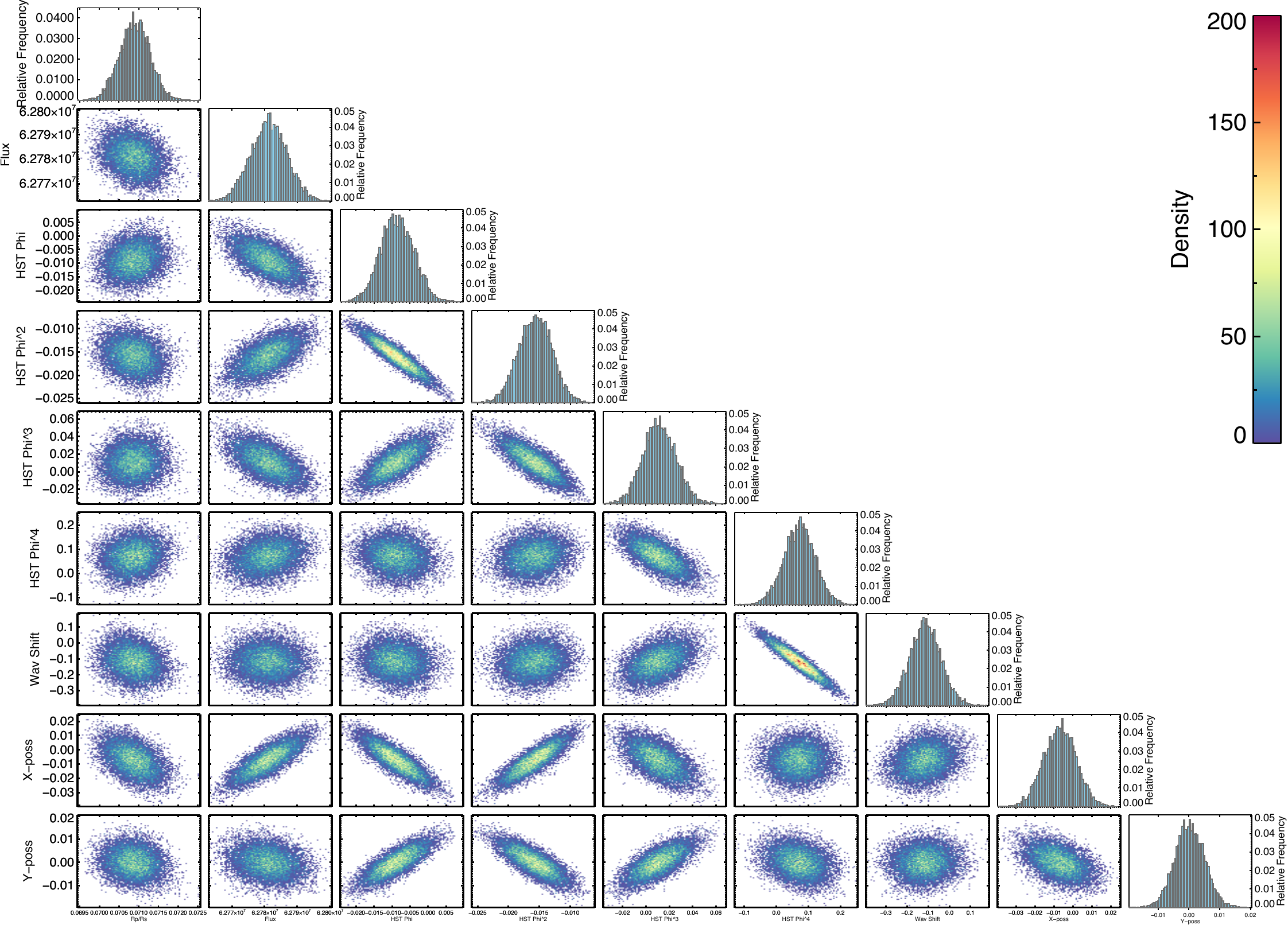}
\caption{Example MCMC posterior distribution using of a transit light curve model
  fit to HST STIS HAT-P-26b data \citep{2017Sci...356..628W}.} 
\label{fig:mcmc}       
\end{figure}

The size of the photometric error bars are a critical, and often
overlooked aspect of light curve fitting.  Typically, the minimum
$\chi^2$ values and best-fit parameters are insensitive to the exact
size of the errors (within reason).  However, the inferred
uncertainties on the best-fit model parameters themselves are dependent upon
the size of the photometric errors/uncertainties given.  In high-precision time-series
photometry, it is often the case that the formal error bars are
dominated by photon noise.  However, for many datasets photon noise is
rarely achieved and can even be factors of a few away from achieving
those levels of precision.  A common practice is to adopt photon-noise
error bars when performing model selection, such that a given model
$f(t)$ performance can be compared to the theoretical limit of the
data, and 
a comparison can be made between the performance of different models.
However, when a satisfactory model is found, the 
photometric error bars themselves are rescaled by the
standard deviation of the residuals, and the model is re-fit.  
Alternatively, for methods such as MCMC that optimise the
log-likelihood, the photometric/white noise level can be fit as a free 
parameter itself. 
Rescaling the error bars in these ways helps
take into account noise sources that are difficult to account for in
standard pipeline reduction routines, and the final uncertainties will
be underestimated if this step (or a similar procedure) is not taken, unless of course photon
noise precisions were achieved. 
In addition, the presence of time-correlated red noise must also be
considered, with the binning technique \citep{2006MNRAS.373..231P} and
wavelets \citep{2009ApJ...704...51C} two recommended methods. 
 With the photometric error
bars set to realistic values, and red noise properly handled, MCMC routines
(e.g. \citealt{2013PASP..125...83E, 2013PASP..125..306F}.) can then be
run to provide the full posterior distribution and marginalised
parameter uncertainties (see Fig. \ref{fig:mcmc}).

After the white light curve is fit, the spectroscopic
channels are then determined and transmission or emission spectra
constructed.  
During this stage, it is typically only the transit or eclipse depth
that is allowed to vary along with the systematics model in the fits.
All other parameters tend to be
fixed to those values adopted for the white light curve, as they are
not expected to vary across spectroscopic channels. This includes the
semi major axis $a$, orbital inclination $i$ and transit/eclipse
mid-time.  An important exception is the limb-darkening coefficients,
which as noted above must be carefully determined for each
spectroscopic channel individually as they are wavelength-dependent.\\

\noindent Below are a few tips to consider regarding transit observations:
\begin{enumerate}
\item{Observe whole time series on the same pixel, sub-pixel if
    possible.}
\item{Choose a detector setting (e.g. gain) to get as many counts as
    reasonably possible per image in order to improve the relative photometric precision.}
\item{Readnoise is almost never important, as we tend to observe bright targets, so the error budget is dominated by photon noise.}
\item{Red noise is always important to consider.}
\item{For MOS, wide slits/apertures are needed to minimise
    differential slit losses between the target and reference stars.}
\item{Have a uniform detector.}
 \item{Be mindful of detector electronics which can introduce noise.}
\item{Get a long enough baseline before and after the transit/eclipse to properly measure the depth.}
\item{As much as possible, have the out-of-transit/eclipse baseline
    sample similar systematics to those in-transit/eclipse.}
 \item{Ingress and egress are not very useful to measure the
     transit/eclipse depth but are needed to
     measure $a$/$R_{star}$, $T_0$, and the inclination.}
 \item{Avoid or correct for non-linearities in the detector.}
 \item{Reference stars need to be as similar as possible to the target
     in type \& magnitude (preferably within 1 mag, and spectral sub-type).}
 \item{Near-IR is hard from the ground, due to terrestrial opacity sources such as water, but not impossible.}
 \item{Philosophy differs in flat-fielding (I suggest gathering 100's, test to see if they improve things or not).}
 \item{Cosmic Rays are important.}
 \item{Visualise your time series data (e.g. with a movie) to see all
     the ways your spectra/photometry changes in time.}
 \item{Be wary of hidden companions that can dilute the flux of the brighter star you're interested in.}
 \item{Be wary of unphysical results.}
\end{enumerate}


\section{Interpreting a Transmission Spectra}
\label{sec:1.4}
In the following section, an analytic transmission spectra formula
is derived from first principles.  Details of this derivation can be found
across \cite{2005ApJ...627L..69F}, \cite{2008A&A...481L..83L}, \cite{2013Sci...342.1473D}, \cite{2017MNRAS.467.2834B} and \cite{2017MNRAS.470.2972H}.  Given that a transmission
spectrum is not a typical astrophysical measurement, the derivation
helps to illustrate how the spectrum depends upon several key
parameters (such as the temperature and molecular abundances), what
quantities can be derived from a transmission spectra, and the nature
of modelling degeneracies. 

\subsection{Analytic Transmission Spectrum Derivation}
\label{sec:1.4.1}

\begin{figure}[t]
  \centering
\includegraphics[scale=.50]{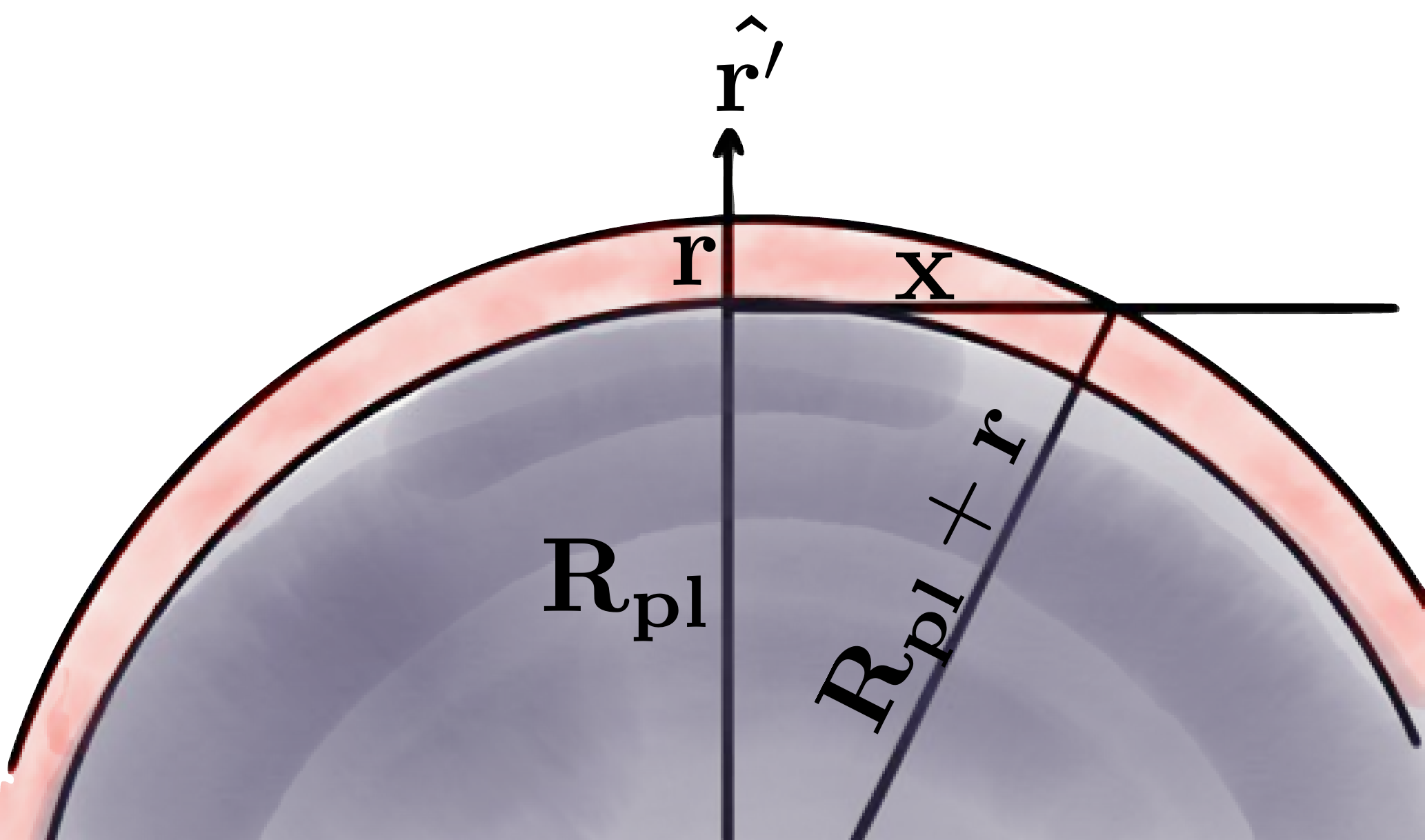}
\caption{Illustration of the transit geometry.  The bulk radius of the planet, $R_{pl}$
  (grey) is illustrated along with an atmospheric layer (red) of
  thickness, $r$.  Stellar light passes in slant transit geometry
  from the terminator through the atmospheric layer a distance $x$.
  The radial coordinate $\hat{r'}$ is also indicated with its origin at
the planet center.} 
\label{fig:geom}       
\end{figure}

The flux blocked during transit including the planet and its atmosphere is given by the ratio of the areas,
\begin{equation}
\frac{\Delta f}{f} =  \frac{\pi R_{pl}^2+A}{\pi R_{star}^2},
\end{equation}
where $A$ is the effective area of the annular region of the
atmosphere observed during transit.  The contribution of the
atmosphere, $A$, is calculated by integrating the absorptivity of the
atmosphere from a reference planetary radius, $R_{pl}$, up to the top of
the atmosphere along the radial coordinate direction $r'$ and is given
by, 
\begin{equation}
A=\int_{R_{pl}}^{\infty} (1-T) 2 \pi r' dr'
\label{eq:A}
\end{equation}
where $T$ is the transmittance which is the fraction of radiation that
is transmitted through a given layer of atmosphere.  The transmittance
is related to the optical depth  $\tau $ using the Beer-Lambert law, 
\begin{equation}
T=e^{-\tau},
\label{eq:Beer}
\end{equation}
and the optical depth in turn can also be written in terms of the cross section of absorbing
species $\sigma_{abs}(\lambda)$, its number density $n_{abs}$, and
integrated along the slant transit geometry (direction $\hat{x}$) ,
\begin{eqnarray}
\label{eq:tau}
\tau=\int_{-\infty}^{+\infty}\sigma_{abs}(\lambda) n_{abs}dx,\\
T=e^{-\int_{-\infty}^{+\infty}\sigma_{abs}(\lambda) n_{abs}dx}.
\end{eqnarray}
As the number density of an atmosphere drops with altitude radially
(direction $\hat{r'}$), a simplifying approximation can be made by
assuming the atmosphere is isothermal with temperature $T$, an ideal
gas with pressure $P=k_B\rho T/\mu$, where $\rho$ is the density, and in hydrostatic
balance,
\begin{equation}
dP=-\rho g dr.
\label{eq:hydro}
\end{equation}
Substituting $\rho$ from the ideal gas law and integrating Eq. \ref{eq:hydro}
then gives the Barometric formula which relates the pressure between
two points (labeled here as 0 and 1) with the altitude difference,
$P_1=P_0e^{-r/H}$, and pressure scale height $H$ or equivalently the number
density between two altitude points,
\begin{equation}
n_1=n_oe^{-r/H}.
\label{eq:numden}
\end{equation}
Typically, the absorbing species in a transmission spectra is a minor
component of the gas, so we can relate the mixing ratio of the
absorbing minor species, $\xi_{abs}$, to the total gas number density, by $n_{abs}=n_o\xi_{abs}$.  
Substituting Eq. \ref{eq:numden} into Eq. \ref{eq:tau} and evaluating
the optical depth at the reference planetary radius and pressure
($R_{pl}$, $P_0$) gives,
\begin{equation}
\tau_0=\xi_{abs}\sigma_{abs}(\lambda) \int_{-\infty}^{+\infty}\frac{P_0}{k_B T}e^{-r'/H}dx,
\label{eq:tauoint}
\end{equation}
where we have made the assumption that the cross section
$\sigma_{abs}(\lambda)$ does not depend upon the atmospheric
pressure.  
The $\hat{x}$ and $\hat{r'}$ coordinates can be related by the Pythagorean theorem (see Fig. \ref{fig:geom}),

\begin{eqnarray}
(R_{pl}+r)^2=R_{pl}^2+x^2\\
x^2=2R_{pl} r + r^2,
\end{eqnarray}
and further simplified assuming planetary radius is much larger than
the atmospheric altitude ($R_{pl}>>r$) which gives
\begin{equation}
r=\frac{x^2}{2R_{pl}}.
\end{equation}
Substituting $r$ into Eq. \ref{eq:tauoint} gives,
\begin{equation}
\tau_0=\xi_{abs}\sigma_{abs}(\lambda) \int_{-\infty}^{+\infty}\frac{P_0}{k_B T}e^{-x^2/2R_{pl}H}dx.
\end{equation}
The integral can be analytically evaluated as,
$\int_{-\infty}^{+\infty} e^{-u^2} du = \sqrt\pi$,
so the reference optical depth becomes,
\begin{equation}
\tau_0=\xi_{abs}\sigma_{abs}(\lambda) \frac{P_0}{k_B T}\sqrt{2 \pi R_{pl}H}.
\label{eq:tauo}
\end{equation}
From Eq. \ref{eq:tauo}, it can be seen that the optical depth depends
on the pressure (i.e. $\tau \propto P$), and again using the
Barometric formula the optical depth at an arbitrary pressure $P$ can
be written as, 
\begin{equation}
\tau=\xi_{abs}\sigma_{abs}(\lambda) \frac{P}{k_B T}\sqrt{2 \pi R_{pl}H}.
\label{eq:tauP}
\end{equation}

Before we can evaluate Eq. \ref{eq:A}, we need a relation between the
the radial coordinate $\hat{r'}$ and optical depths $\tau$ and
$d\tau$.  Using the Barometric formula in terms of pressure, and the
fact that the optical depth is proportional to pressure gives,
\begin{eqnarray}
\frac{\tau}{\tau_0}=\frac{P}{P_0}=e^{-r/H},\\
\tau=\tau_0 e^{-r/H},
\end{eqnarray}
the altitude difference, $r$, is then related to the reference radius
of the planet and radial coordinate ${r'}$ as $r=r'-R_{pl}$, thus 
\begin{eqnarray}
\tau=\tau_0 e^{-(r'-R_{pl})/H},\\
\ln\left (\frac{\tau}{\tau_0}\right) = -\frac{r'}{H}+\frac{R_{pl}}{H},\\
r'=R_{pl}-H\ln\frac{\tau}{\tau_0},
\end{eqnarray}
and
\begin{equation}
dr'=-\frac{H}{\tau}d\tau.
\label{eq:drtau}
\end{equation}
Substituting these values $r'$ of $dr'$ into Eq. \ref{eq:A} along with
Eq. \ref{eq:tauP} and Eq. \ref{eq:Beer} then gives,
\begin{equation}
A=2 \pi \int_{\tau_0}^{0} (1-e^{-\tau}) \left(R_{pl}-H\ln\frac{\tau}{\tau_0} \right) \left(-\frac{H}{\tau}d\tau \right),
\label{eq:Atau}
\end{equation}
where the integration limits have also been converted into the
appropriate optical depth limits.
After a slight rearrangement, Eq. \ref{eq:Atau} can be written
\begin{equation}
A=2 \pi H R_{pl} \int_{0}^{\tau_0} \left(\frac{1-e^{-\tau}}{\tau}\right) \left(1+\frac{H}{R_{pl}}\ln\frac{\tau_0}{\tau} \right) d\tau,
\end{equation}
which can then be further simplified assuming $H/R_{pl}<<1$ so a term can be
neglected to give,
\begin{equation}
A=2 \pi H R_{pl} \int_{0}^{\tau_0} \left(\frac{1-e^{-\tau}}{\tau}\right) d\tau.
\label{eq:A4}
\end{equation}
The integral can be found in \cite{1960ratr.book.....C},
\begin{equation}
E_1=-\gamma-\ln\tau_0+\int_{0}^{\tau_0}\left(\frac{1-e^{-\tau}}{\tau}\right)d\tau,
\end{equation}
where $\gamma$ is the Euler-Mascheroni constant
($\gamma=$0.577215664901...) and $E_1$ is an exponential integral
which is a transcendental function.  Thus, Eq. \ref{eq:A4} becomes,
\begin{equation}
A=2 \pi H R_{pl} [E_1+\gamma+\ln\tau_0].
\label{eq:A5}
\end{equation}
During a transit, we measure the transit-depth radius of the combined
planet and atmosphere.  Thus to relate \ref{eq:A5} to the transit 
radius, we can set the measured transit depth
altitude $r_{eq}$ so it produces the same absorption depth as the planet with
its translucent atmosphere occulting area, $A$, giving,
\begin{eqnarray}
A=2 \pi R_{pl} r_{eq} = 2 \pi H R_{pl} [E_1+\gamma+\ln\tau_0],\\
r_{eq} = H [E_1+\gamma+\ln\tau_0].
\end{eqnarray}
As $\tau_0$ is the optical depth an arbitrary reference
pressure-altitude level, we can set this value to be very large ($\tau_0>>1$) to use the limit
where  $E_1 \rightarrow 0$ as $\tau_0 \rightarrow \infty$, 
\begin{equation}
r_{eq} = H [\gamma+\ln\tau_0].
\label{eq:req1}
\end{equation}
Eq. \ref{eq:req1} can then be rearranged and the radius converted to optical
depth (Eq. \ref{eq:drtau}) to figure out the equivalent optical depth $\tau_{eq}$ where the
transmission spectrum becomes optically thick in slant transit geometry and
corresponds to the measured transit-depth radius giving,
\begin{eqnarray}
r_{eq} /H= \gamma+\ln\tau_0,\\
e^{-r_{eq} /H}={\tau_{eq}/\tau_0}=e^{- \gamma-\ln\tau_0},\\
\tau_{eq}/\tau_0=e^{-\gamma}/\tau_0,\\
\tau_{eq}=e^{-\gamma}=0.561459.
\label{eq:taueq}
\end{eqnarray}
\cite{2008A&A...481L..83L} first derived $\tau_{eq}$ numerically, where
$\tau_{eq}=0.56$ is seen to be an accurate approximation (given the
terms we have neglected) for most planetary
atmospheres as long as (30$>R_{pl}/H>$300).
Finally, we can substitute Eq. \ref{eq:taueq} and Eq. \ref{eq:tauo}
into Eq. 
\ref{eq:req1} to give,
\begin{eqnarray}
r_{eq} = H \left[-\ln\tau_{eq}+\ln \left( \xi_{abs}\sigma_{abs}(\lambda) \frac{P_0}{k_B T}\sqrt{2 \pi R_{pl}H} \right) \right],\\
r_{eq} = H \left[\ln \left(  \frac{\xi_{abs}\sigma_{abs}(\lambda)P_0}{\tau_{eq}}\sqrt{\frac{2 \pi R_{pl}H}{k_B^2 T^2}} \right) \right],
\end{eqnarray}
and $r_{eq}$ relabelled as $z(\lambda)$ which then derives the \cite{2008A&A...481L..83L} transmission spectrum formula, 
\begin{equation}
z(\lambda)= H \ln \left(  \frac{\xi_{abs}\sigma_{abs}(\lambda)P_0}{\tau_{eq}}\sqrt{\frac{2 \pi R_{pl}}{k_B T \mu g}} \right).
\label{eq:Lec08}
\end{equation}

\subsection{Analytic Transmission Spectrum Applications}
\label{sec:1.4.2}
The first application of Eq. \ref{eq:Lec08} was to interpret the
transmission spectrum of HD~189733b from \cite{2008MNRAS.385..109P} in which
atmospheric haze on an exoplanet was first discovered.  If the cross
section $\sigma_{abs}(\lambda)$ is known, then the altitude difference
between two wavelengths, $dz$, allows the pressure scale height $H$ to be directly
measured, which directly leads from Eq. \ref{eq:Lec08} to,
\begin{equation}
T= \frac{\mu g}{k_B} \left(\frac{d\ln\sigma}{d\lambda} \right)^{-1} \frac{dz(\lambda)}{d\lambda}.
\end{equation}
This temperature measurement will be accurate in the case where the
absorbing species in the transmission spectra has been robustly identified, such as is often the case
for Na, K, or H$_2$O, and when the mean molecular weight of the atmosphere is also
known.  Thus, the terminator temperatures in hot Jupiters can often be accurately
measured, given the atmosphere is H/He dominated; though that may not
be the case for super-Earths which could have much heavier non-H/He secondary atmospheres.

For HD~189733b, the slope of the transmission spectra indicated a
scattering slope, and the cross section could be assumed to follow a
power law of index $\alpha$, such that
$\sigma=\sigma_0(\lambda/\lambda_0)^{\alpha}$.  In this case, 
\begin{equation}
\alpha T = \frac{\mu g}{k_B}  \frac{dR_{pl}(\lambda)}{d\ln\lambda},
\end{equation}
making the transmission spectrum slope proportional to the product
$\alpha T$.  In the case of pure Rayleigh scattering, $\alpha$=-4 and
the temperature can be derived, though in general if the pure Rayleigh
scattering is not apparent, then the constrained quantity will be
$\alpha T$ and a degeneracy will exist between the power law index
and the atmospheric temperature.

Eq. \ref{eq:Lec08} can be used to straightforwardly make an entire optical
transmission spectrum for a hot Jupiter, given that theoretical models
have shown a typical hot Jupiter ($\sim$1200 K) will be dominated by
Na, K, and Rayleigh scattering \citep{2000ApJ...537..916S,
  2001ApJ...560..413H, 2001ApJ...553.1006B}.  Na and K are both doublets, so by including only four
absorption lines and a scattering component, the majority of an hot Jupiter
optical transmission spectrum can be modelled.

For largely clear atmospheres, Na and K can both exitbit large
pressure-broadened wings which will dominate the optical opacity.  
These wings can be calculated analytically using a Voigt line profile, $H(a,u)$,
and statistical theory, which predicts the collision-broadened alkali line
shapes will vary with frequency $\nu$ as $(\nu-\nu_0)^{-3/2}$ outside of an impact
region, which lies between the line centre frequency $\nu_{0}$
and a detuning frequency $\Delta \sigma$ away from $\nu_{0}$.

Within the impact region, 
the cross-sections for both the sodium and potassium doublets 
can then calculated following \cite{2000ApJ...531..438B} and \cite{2005A&A...436..719I} as,
\begin{equation} 
\sigma(\lambda)=\frac{\pi e^2}{m_e c}\frac{f}{\Delta\nu_D \sqrt{\pi}}H(a,u),
\end{equation}
where $f$ is the absorption oscillator strength of the spectral line,
$m_e$ is the mass of the electron, $e$ the electron charge.  The Voigt
profile $H(a,u)$ is defined in terms of the
Voigt damping parameter $a$ and a frequency offset $u$.
The frequency offset is calculated as 
$u=(\nu-\nu_0)/\Delta\nu_D$, where $\Delta\nu_D$ is the Doppler width 
given by $\Delta \nu_{\rm D} = \nu_{0}/c \sqrt{2kT/\mu_{\rm Na,K}}$,
with $c$ the speed of light and $\mu_{\rm Na,K}$ the mean molecular
weight of sodium or potassium.
The damping parameter is given by $a=\Gamma / (4\pi\Delta\nu_{\rm D})$,
where the transition rate $\Gamma$ is calculated following
\begin{equation}  
\label{Eq:gamma}
\Gamma=\gamma+\Gamma_{\rm col},
\end{equation}
where $\gamma$ is the spontaneous decay rate and $\Gamma_{\rm col}$ is the
half-width calculated from classical impact theory.  Assuming a Van der
Waals force gives,
\begin{equation}  
\Gamma_{\rm col}=0.071(T/2000)^{-0.7}~\rm{cm}^{-1} \rm{atm}^{-1}
\end{equation}
for Na and
\begin{equation}  
\Gamma_{\rm col}=0.14(T/2000)^{-0.7}~\rm{cm}^{-1} \rm{atm}^{-1} 
\end{equation}
for K \citep{2000ApJ...531..438B,2005A&A...436..719I}.  

Outside of the impact region, the $(\nu-\nu_0)^{-3/2}$ power-law line shape is truncated using an exponential cutoff term of the form
$e^{-qh(\nu-\nu_0)/kT}$, where $h$ is Planck's constant and $q$ a parameter of order unity, to prevent the line wing opacity from becoming
overly large at large frequency separations. 
The detuning frequency, $\Delta \sigma$, can be estimated from
\cite{2000ApJ...531..438B} using 
\begin{equation}  
\Delta \sigma = 30(T/500~{\rm K})^{0.6}~{\rm cm}^{-1}.
\end{equation}
for the sodium doublet and
\begin{equation}  
\Delta \sigma = 20(T/500~{\rm K})^{0.6}~{\rm cm}^{-1}.
\end{equation}
for the potassium doublet.

The wavelength-dependent total cross sections of the sodium or
potassium D1 and D2 doublet can then be summed together,
$\sigma(\lambda)_{Na,K}=\sigma(\lambda)_{D1}+\sigma(\lambda)_{D2}$, and the
sodium and potassium opacities can also be summed together along with
their abundances into Eq. \ref{eq:Lec08} as,
\begin{equation}  
\xi_{abs}\sigma_{abs}=\xi_{Na}\sigma_{Na}+\xi_{K}\sigma_{K}.
\end{equation}
Finally, a Rayleigh scattering component can also be added to the
alkali lines, given that
$\sigma_0$ = $2.52 \times 10^{-28}$ $cm^2$ at $\lambda_0=$750 nm for
molecular hydrogen and for a hot Jupiter $\xi_{H_2}\sim1$.

\begin{figure}[t]
  \centering
\includegraphics[scale=.54]{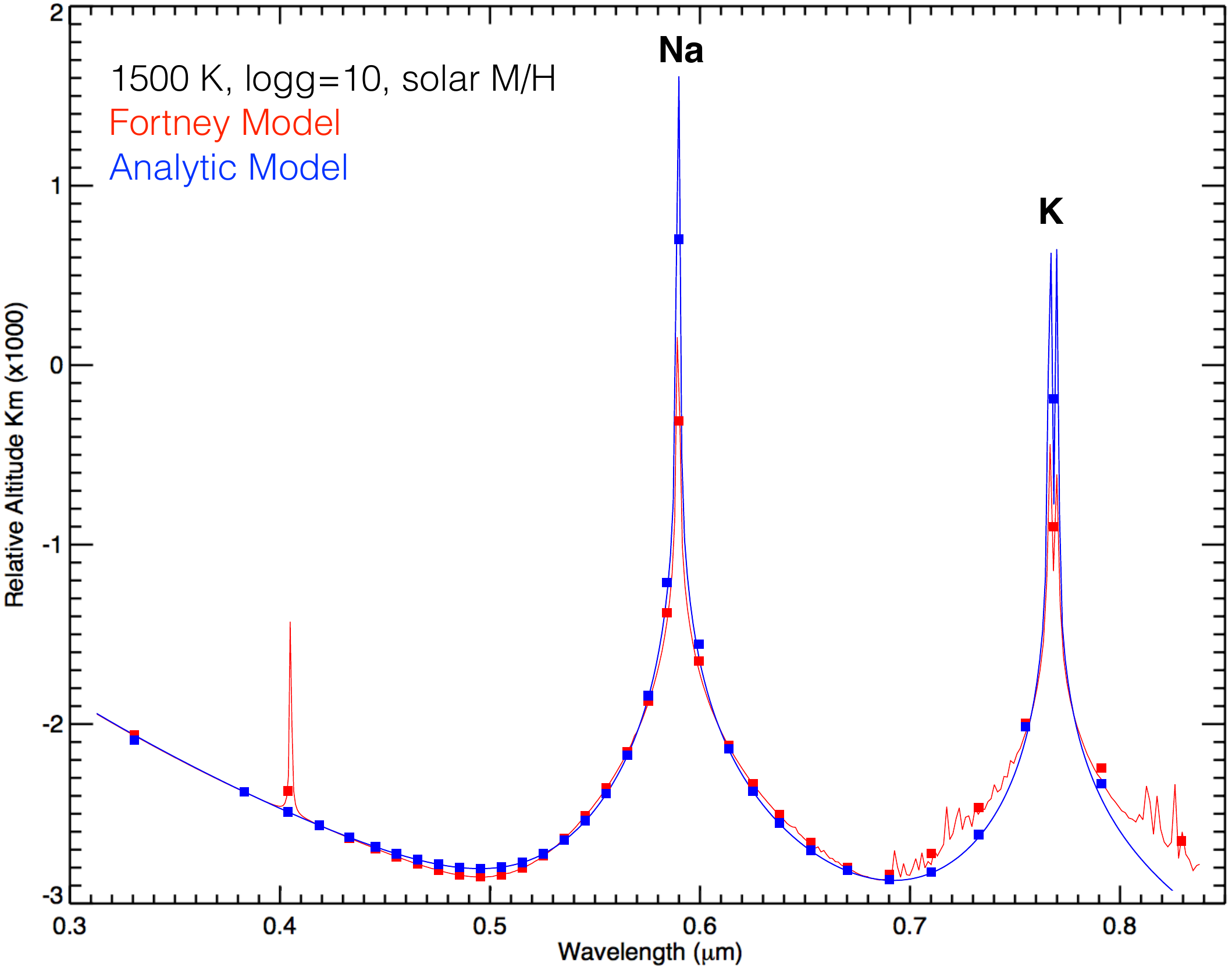}
\caption{Comparison between an analytic transmission spectrum model
  for a 1500 K hot Jupiter calculated using Eq. \ref{eq:Lec08} (blue)
  and a model from \cite{2010ApJ...709.1396F}.} 
\label{fig:LecFortCompare}       
\end{figure}

A comparison between the analytic transmission spectrum (Eq. \ref{eq:Lec08})
and a numerical model from \cite{2010ApJ...709.1396F} is shown in Figure
\ref{fig:LecFortCompare}.  The analytic model reproduces the Na and K
lines profiles well, and certainly better than the accuracy of any
transmission spectral data to date.  The small differences seen are mainly due to the inclusion of
molecules such as H$_2$O in the \cite{2010ApJ...709.1396F} model, which can be seen as weak lines near the K
doublet, and the inclusion of weaker Na/K lines (e.g 0.4 $\mu m$) which have not
been added in this example analytic model.  

\begin{figure}[t]
  \centering
\includegraphics[scale=.54]{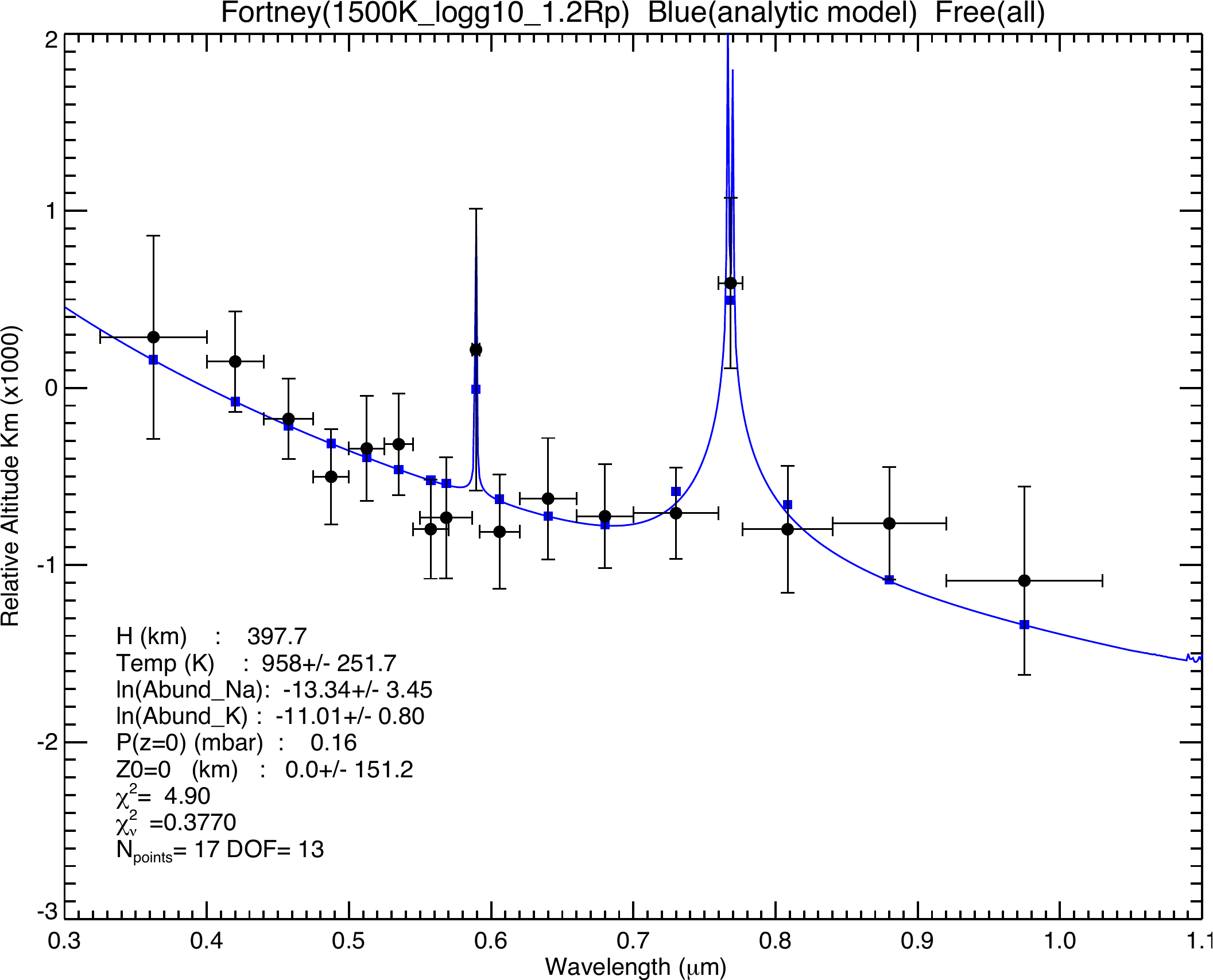}
\caption{Analytic atmospheric model fit of the WASP-6b HST
  transmission spectrum from
  \cite{2015MNRAS.447..463N}.  Four parameters are fit, the
  temperature, Na and K abundances, and a baseline reference planet radius.  The best-fit parameters and $\chi^2$
  statistics are indicated on the plot.} 
\label{fig:W6model}       
\end{figure}

The analytic transmission spectrum (Eq. \ref{eq:Lec08}) can easily be
used to fit data in a retrieval model exercise (e.g. \citealt{2015MNRAS.446.2428S}), which given the few
parameters and analytic nature is very fast making it highly conducive
to Markov chain Monte Carlo techniques (MCMC).  In Fig. \ref{fig:W6model}
the WASP-6b optical HST data from \cite{2015MNRAS.447..463N} is well fit using only
four parameters: the abundances of Na and K, as well as the temperature and
baseline planetary radius.

In general, molecular species can also be modeled with the analytic
transmission spectrum (Eq. \ref{eq:Lec08}) as well, in this case the number
of spectral lines jumps from four up to 10$^9$ or 10$^{10}$ depending
on the line list and species in question, which dramatically increases the
computational burden.  Nevertheless, even if one uses a fully
numerical tool to calculate and model transmission spectra, it is a
good idea to keep Eq. \ref{eq:Lec08} in mind such that a physical
intuition can be preserved.

A further consequence of Eq. \ref{eq:Lec08}  is that abundance ratios
can be very precisely measured in a transmission spectrum.  This
follows from measuring the altitude difference in the spectra between
the two (or more) wavelengths where the two species dominate.  For
example, taking the difference in altitudes between the transmission
spectra at the Na core wavelengths, $z_{Na}$, and potassium core wavelengths $z_{K}$,
\begin{equation}
z_{Na}-z_{K}= H \ln \left( \frac{\xi_{Na}\sigma_{Na}(\lambda)P_0}{\tau_{eq}}\sqrt{\frac{2 \pi R_{pl}}{k_B T \mu g}} \right) - H \ln \left(  \frac{\xi_{K}\sigma_{K}(\lambda)P_0}{\tau_{eq}}\sqrt{\frac{2 \pi R_{pl}}{k_B T \mu g}} \right),
\end{equation}
which can be simplified to,
\begin{eqnarray}
{\frac{z_{Na}-z_{K}}{H}}= \ln \left(  \frac{\xi_{Na}\sigma_{Na}(\lambda)P_0}{\tau_{eq}} \right) - \ln \left(  \frac{\xi_{K}\sigma_{K}(\lambda)P_0}{\tau_{eq}} \right),\\
{\frac{z_{Na}-z_{K}}{H}}= \ln \left(  \xi_{Na}\sigma_{Na}(\lambda)\right) - \ln \left(  \xi_{K}\sigma_{K}(\lambda) \right),\\
e^{\frac{z_{Na}-z_{K}}{H}}=\frac{\xi_{Na}\sigma_{Na}(\lambda)}{\xi_{K}\sigma_{K}(\lambda)},\\
\frac{\xi_{Na} } { \xi_{K} } = \frac {\sigma_{K}} { \sigma_{Na}} \exp\left({ \frac{z_{Na}-z_{K}} {H} }\right).
\end{eqnarray}
Almost all of the constants cancel, including importantly $P_0$ and
$R_{pl}$, and the measured transit difference between the wavelengths of
the two species directly translates into the abundance ratios between
the two species.  In the near-future, this aspect of transmission
spectroscopy should lead to well measured C/O ratios in planetary
atmospheres given the ratios of the dominant species (H$_2$O, CO$_2$,
CO, \& CH$_4$) may all be determined, e.g.
\begin{equation}
\frac{\xi_{H_2O} } { \xi_{CO} } = \frac {\sigma_{CO}} { \sigma_{H_2O}} \exp\left({ \frac{z_{H_2O}-z_{CO}} {H} }\right).
\end{equation}
with early attempts of this measurement found in \cite{2009ApJ...699..478D}
using Spitzer data.

\subsection{Transmission Spectrum Degeneracies}
\label{sec:1.4.3}
An important degeneracy is apparent from Eq. \ref{eq:Lec08} as the
transmission spectrum is determined by the quantity of $\xi_{abs}P_0$.  In general, the baseline pressure $P_0$ at the
reference planetary radius is not known. Thus, it is typically difficult to measure
absolute abundances in a transmission spectrum, which will be limited by the degeneracy
between the abundance and baseline pressure.  The degeneracy can
also be re-cast as a degeneracy between $\xi_{abs}$ and the reference
planetary radius $R_{pl}$.  Further discussion of degeneracies  can be found in \cite{2012ApJ...753..100B} and \cite{2017MNRAS.470.2972H}.

One way to lift the  $\xi_{abs}-P_0$ degeneracy is to identify
Rayleigh scattering by molecular H$_2$.  Models such as those shown in
Fig. \ref{fig:LecFortCompare}  predict for clear hot Jupiter atmospheres that
short-ward of about 0.5 $\mu$m, the H$_2$ molecular scattering should
dominate the opacity.  If this is the case, then the abundance of H$_2$
can be well approximated given  $\xi_{H_2}\sim1$, and the reference
pressure $P_0$ at $z=0$ can be determined.  From Eq. \ref{eq:Lec08}, in the
H$_2$-Rayleigh spectral region,
the pressure P$_0$ at an altitude corresponding to the radius at wavelength
$\lambda_0$ is
\begin{equation}
P_0= \frac{ \tau_{eq} }{ \sigma_{0} } \sqrt{ \frac{k_B T \mu g}{2 \pi R_{pl}}},
\end{equation}
where $\sigma_{0}$ is the Rayleigh scattering cross section at
$\lambda_0$.  If $P_0$ can be determined from this method, the
absolute abundances of all species identified in a transmission spectra
(including molecular species at longer infrared wavelengths) can then
be determined.  In practice, hazes and clouds have often been
observed to mask the H$_2$ Rayleigh scattering signatures.  However,
even if this is the case the short-wavelength region is still an
important wavelength region to measure, as this data can constrain
the cloud properties and rule out large classes of models, thereby
constraining parameter space and limiting the $\xi_{abs}-P_0$ degeneracy. 

In addition to the $\xi_{abs}-P_0$ degeneracy, the transmission
spectra is also seen from Eq. \ref{eq:Lec08} to scale (to first order) with the
pressure scale height $H=k_BT/\mu g$.  The $H$ itself can typically be well
determined from a well-measured transmission spectra.  For hot
Jupiter exoplanets which also have a measured mass, the surface
gravity $g$ is also known and $\mu$ can safely be assumed to be
dominated by a H/He mixture leaving just the temperature as the
unknown quantity that can be measured from the transmission spectrum.  However,
for non-gas giant exoplanets the molecular weight of the atmosphere
can be unknown to perhaps an order of magnitude or more given
atmospheres dominated for example by N$_2$, H$_2$O, or CO$_2$ are
feasible.  In addition, it is also more challenging to measure the mass
of small exoplanets via the radial velocity method, which can also
hinder a precise determination of $g$.  Thus, in these cases there
will be a large $T-g-\mu$ degeneracy, and the individual quantities
will be difficult to constrain with the transmission spectrum alone.



\begin{acknowledgement}
David K. Sing acknowledges the support from the meeting organisers for
travel and accommodation at the meeting.  DKS also gives thanks to
Tom Evans, Nikolay Nikolov, Kevin Stevenson, and Hannah Wakeford for the use
of several figures.  DKS gives further thanks to Aarynn Carter, Tom Evans, Nikolay
Nikolov, Jayesh Goyal,  Jessica Spake, and Hannah Wakeford for
reviewing the manuscript. 
DKS acknowledges funding from the European Research Council under
the European Union’s Seventh Framework Programme (FP7/2007-2013)/ERC grant agreement number 336792.
\end{acknowledgement}

\bibliographystyle{yahapj} 
\bibliography{ASES_Sing} 

\end{document}